\documentclass[journal=jctcce,manuscript=article]{achemso}
\setkeys{acs}{articletitle = true}

\usepackage{amsmath}
\usepackage{amssymb}
\usepackage{amsfonts}
\usepackage{color}
\usepackage{bm}
\usepackage[english]{babel}
\usepackage{hyperref}
\usepackage{siunitx}
\usepackage{graphicx}
\usepackage{enumerate}
\usepackage[version=3]{mhchem} 
\usepackage{physics}
\usepackage{setspace}
\usepackage{geometry}
\usepackage[utf8]{inputenc}
\usepackage{polski}
\newcommand{\onlinecite}[1]{\hspace{-1 ex} \nocite{#1}\citenum{#1}}
 
\newcommand{\kB}{\ensuremath{k_{\rm B}}}
\newcommand{\kb}{\ensuremath{\kB}}
\newcommand{\kT}{\ensuremath{\kb T}}

\newcommand{\widefigurewidth}{0.75\textwidth}

\author{Jiale Shi}
\affiliation{%
  Department of Chemical and Biomolecular Engineering, %
  University of Notre Dame, %
  Notre Dame, Indiana 46556, United States
}

\author{Fahed Albreiki}%
\affiliation{%
Department of Chemical and Biomolecular Engineering, University of California, Los Angeles, Los Angeles, California 90095, United States
}

\author{Yamil J. Col\'{o}n }%
\affiliation{%
  Department of Chemical and Biomolecular Engineering, %
  University of Notre Dame, %
  Notre Dame, Indiana 46556, United States
}

\author{Samanvaya Srivastava}%
\affiliation{%
Department of Chemical and Biomolecular Engineering, University of California, Los Angeles, Los Angeles, California 90095, United States
}
\alsoaffiliation{%
California NanoSystems Institute, Center for Biological
Physic, University of California, Los Angeles, Los Angeles, California 90095, United States
}%
\alsoaffiliation{%
 Institute for Carbon Management, University of California, Los Angeles, Los Angeles, California 90095, United States
}%
\alsoaffiliation{%
 Center for Biological Physics, University of California, Los Angeles, Los Angeles, California 90095, United States
}%

\author{Jonathan K. Whitmer}
\email{jwhitme1@nd.edu}
\affiliation{%
  Department of Chemical and Biomolecular Engineering, %
  University of Notre Dame, %
  Notre Dame, Indiana 46556, United States
}
\alsoaffiliation{%
Department of Chemistry and Biochemistry,% 
University of Notre Dame, Notre Dame, %
Indiana 46556, United States
}%

\title[]{Transfer Learning Facilitates the Prediction of Polymer--Surface Adhesion Strength}

\begin{document}

\begin{abstract}
Machine learning (ML) accelerates the exploration of material properties and their links to the structure of the underlying molecules. In previous work [J. Shi, M. J. Quevillon, P. H. A. Valença, and J. K. Whitmer, \textit{ACS Appl. Mater. Interfaces.}, 2022, 14, 32, 37161–37169 ], ML models were applied to predict the adhesive free energy of polymer--surface interactions with high accuracy from the knowledge of the sequence data, demonstrating successes in inverse-design of polymer sequence for known surface compositions. While the method was shown to be successful in designing polymers for a known surface, extensive datasets were needed for each specific surface in order to train the surrogate models. Ideally, one should be able to infer information about similar surfaces without having to regenerate a full complement of adhesion data for each new case. In the current work, we demonstrate a transfer learning (TL) technique using a deep neural network to improve the accuracy of ML models trained on small datasets by pre-training on a larger database from a related system and fine-tuning the weights of all layers with a small amount of additional data. The shared knowledge from the pre-trained model facilitates the prediction accuracy significantly on small datasets. We also explore the limits of database size on accuracy and the optimal tuning of network architecture and parameters for our learning tasks. While applied to a relatively simple coarse-grained (CG) polymer model, the general lessons of this study apply to detailed modeling studies and the broader problems of inverse materials design.
\end{abstract}
\maketitle

\section*{Introduction}
Numerous industrial applications and biological phenomena involve chemically specific polymer--surface interactions, from  ink absorption on paper,~\cite{chakraborty2001polymer,li2019advanced} and semiconductor fabrication and coating,~\cite{kim2003epitaxial,ikawa2012simple}  to the design and synthesis of artificial tissues~\cite{Annabi2014} and viruses recognizing receptors on a cell surface.~\cite{xiu2020Inhibitors,wong2021sars,callaway2020making, plante2021spike, hie2021learning} The use of highly tuned sequence-defined polymers is attractive in controlling phase behavior, stabilizing interfaces, and promoting adhesion. Sequence-dependent adsorption of polymers to patterned surfaces has been studied through traditional theoretical and computational approaches~\cite{ozboyaci2016modeling, chakraborty1998simple, muthukumar1995pattern,chakraborty2001disordered, Muthukumar1998pattern, Chauhan2021crowding, kriksin2005adsorption, tereshkin2022predicting} and machine learning methods,~\cite{shi2021predicting} emphasizing the importance of polymer sequence in determining the adsorption energies.~\cite{kriksin2005adsorption,shi2021predicting} 

Machine learning (ML) and artificial intelligence (AI)~\cite{bishop2006PRML,murphy2012machine,pml1Book,pml2Book,artrith2021best, de2019new, gormley2021machine, huang2020deepPurpose,liang2022machine,briceno2022pemnet,sattari2021data,Sevgen2020combined,sidky2018learning} have achieved dramatic success in determining the behaviors and properties of polymer and biomacromolecule systems,~\cite{alphafold2, tunyasuvunakool2021highly, RoseTTAFold, Webb2020Target, statt2020model, statt2021unsupervised, Meenakshisundaram2017design, ma2018determining, arora2021random} including predicting protein structure,~\cite{alphafold2,tunyasuvunakool2021highly,RoseTTAFold} polymer structures (such as radius of gyration in solvent),~\cite{Webb2020Target,patel2022featurization} and thermodynamic properties (such as polymer glass transition temperature, $T_g$).~\cite{ma2018determining,ma2022exploring,ma2019evaluating}  However, the wide-ranging chemical sequence, topological space, and mass distribution of the polymer are too extensive to explore.~\cite{lin2019bigsmiles,patel2022featurization} For example, even for linear binary copolymers with twenty monomers, the number of possible sequences is approximately one million. The chemical space becomes exponentially large if more monomer types, variable degrees of polymerization, non-uniform topologies, and mass distributions enter the description. ML techniques can help, but often provide knowledge highly specific to the immediate problem and require significant new datasets to incorporate information outside the original scope. For example, our prior work (see Ref.~\onlinecite{shi2021predicting}) utilized ML models to predict the adhesive free energy of polymer--surface interactions with high accuracy and aid the inverse-design of polymer sequence for known surface compositions, but exploring adhesion of such a polymer to a substrate requires about 8000 data points to train an accurate ML model for each decorated surface. Often, ML models are inaccurate or overfit when trained on small datasets. At the same time, in both industrial applications and biological settings, the surface patterns vary substantially, both structurally and randomly. Collecting large datasets for every patterned surface from thousands or millions of new experiments or simulations is, therefore, prohibitively difficult and expensive. In realistic situations,  it may only be feasible to collect tens to hundreds of new data points. Data-driven ML modeling is easier to implement but often necessitates large datasets that could be difficult to obtain.~\cite{WangTLTutorial2018, yang_zhang_dai_pan_2020, NIPS2014_375c7134} Therefore, our aim here is to determine the minimum amount of additional computation necessary to obtain an accurate binding model, building as much as possible on prior knowledge. 

Transfer learning (TL) can be a valuable technique to overcome the dilemma of insufficient data.~\cite{WangTLTutorial2018,yang_zhang_dai_pan_2020,NIPS2014_375c7134} In TL, an ML model initially pre-trained for a given task on a large dataset of the source domain is utilized as the base to train a model for a new task by fine-tuning a small dataset of the target domain.~\cite{WangTLTutorial2018,yang_zhang_dai_pan_2020,NIPS2014_375c7134,briceno2022pemnet,zhuang2020comprehensive} Typically, TL can improve the model's accuracy if the source and target domains are closely related.~\cite{WangTLTutorial2018,yang_zhang_dai_pan_2020,NIPS2014_375c7134,briceno2022pemnet,NIPS2014_375c7134,pan2009survey} TL has achieved considerable success in speech recognition,~\cite{wang2015transfer,kunze2017transfer} image recognition,~\cite{ng2015deep,yang2021image} and natural language processing.~\cite{gpt3,bert} In addition, TL has also been successfully utilized in materials informatics studies~\cite{Tsubaki2021quantum,Sultan2018transferable,kaser2021transfer} such as structural prediction of gas adsorption in MOFs,~\cite{ma2020transfer} phonon properties in semiconductors,~\cite{liu2020leverage} and thermal conductivity~\cite{wu2019machine}  and electrochemical properties~\cite{briceno2022pemnet} of polymers. However, these studies typically do not explore the explicit inverse design problem involved in materials design: what molecular structures, subject to reasonable constraints, are best for a given application?

In this study, we demonstrate the ability of transfer learning to leverage the prediction performance of adhesive free energies between polymer chains with a defined sequence and patterned surfaces via fine-tuning a pre-trained model. The source domain and learning task come from a large dataset of polymer-surface interactions with one patterned surface.~\cite{shi2021predicting} The target domain and learning task come from a small dataset of polymer-surface interactions with a different patterned surface.~\cite{shi2021predicting} We utilize a deep neural network architecture to perform transfer learning and characterize the improvements on three example cases. We also explore the limits of database size on accuracy and the optimal tuning of network architecture and parameters for our learning tasks.

\section*{Methods}
\label{sec:TL_methods}

\subsection*{Data Set} 
\begin{figure}[h]
	\begin{center}
	\includegraphics[width=\widefigurewidth]{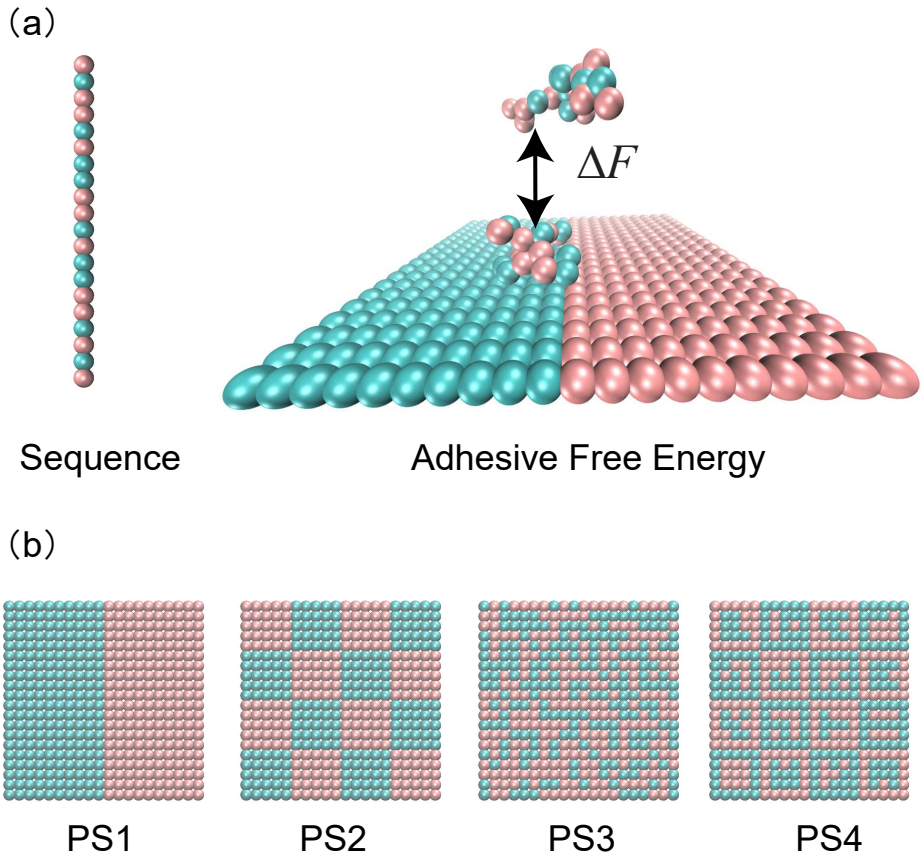}
	\end{center}
	\caption{A schematic of the data sets about adhesive free energies of sequence defined polymers with patterned surfaces from the work of Shi, \textit{et al}.~\cite{shi2021predicting}
	(a) Every data point includes: A sequence-defined polymer and its adhesive free energy with a patterned surface. Each dataset contains $2\times 10^4$ sequence defined polymers and their adhesive free energies with one patterned surface.
	Therefore, for simplification, we use the name of the patterned surface to represent each data set.
	(b) There are four such datasets (Data-PS1, Data-PS2, Data-PS3 and Data-PS4) for four different patterned surfaces( PS1, PS2, PS3, and PS4). 
	}
	\label{fig:data}
\end{figure}

The data sets used in this work  are from our recent work, Shi, \textit{et al.} (Ref.~\onlinecite{shi2021predicting}). As shown in Figure~\ref{fig:data} (a), every data point includes one sequence-defined polymer and its adhesive free energy $\Delta F$ with a patterned surface. The $\Delta F$ were generated by LAMMPS molecular dynamic simulations~\cite{LAMMPS} coupled with adaptive biasing force (ABF) method~\cite{Darve2008adaptive}  SSAGES.\cite{SSAGES, shi2022free, shi2020automated, huang2020deepPurpose, leonhard2019accurate, cortes2021a} The polymer chain and surface are both composed of two types of beads, denoted "red" beads and "green" by their visualization in Figure~\ref{fig:data}. The polymer is modeled as a flexible 20-bead linear chain. The surface is holonomically constrained, with a simple square lattice of beads having dimension $20 \sigma \times 20 \sigma$ for a total of 400 beads. Each dataset contains $2\times 10^4$ sequence-defined polymers and their adhesive free energies with one patterned surface. There are four different data sets, one for each pattern shown in Figure ~\ref{fig:data} (b): PS1, which is composed of half red beads and half green beads in two stripes. $N_{\rm red} = 200$ and $N_{\rm green} = 200$; PS2, which is composed of 16 alternate small size squares ($5\sigma \times 5\sigma$) of red and green beads with the same overall composition as PS1; PS3, where each bead was randomly generated with a probability of 0.5 for each site to be red or green resulting in  $N_{\rm red} = 184$ and $N_{\rm green} = 216$; and PS4, which is built upon PS2, but randomized within the interior of the $5\sigma \times 5\sigma$ squares resulting in a total of $N_{\rm red} = 206$ and $N_{\rm green} = 194$. PS3 and PS4 allow exploration of the role of randomizing effects on our adhesive models, with PS4 including randomness within an overall structure rather than only randomness. For simplicity, we use the name of the patterned surface to represent each data set, called Data-PS1, Data-PS2, Data-PS3, and Data-PS4. Detailed distributions and analysis of the adhesive free energy datasets are available in Ref.~\onlinecite{shi2021predicting}; reduced metrics corresponding to Gaussian fit paramters for each free energy distribution Data-PS1, Data-PS2, Data-PS3, and Data-PS4 are shown in Table~\ref{tab:datadistribution}. Additional details for generating the datasets are discussed in the previous work\cite{shi2021predicting}. All datasets are available online at \url{https://github.com/shijiale0609/ML_PSI}.

\begin{center}
\begin{table}
\begin{tabular}{|c|c|c|c|}
\hline
Dataset & $\mu (\kT)$  & $\sigma  (\kT)$\\
\hline
Data-PS1 & 15.66 &  2.89 \\
\hline
Data-PS2 & 13.84 & 1.55 \\
\hline
Data-PS3 & 8.96 & 0.77  \\
\hline
Data-PS4 & 8.20  & 0.31  \\
\hline
\end{tabular}
\caption{\label{tab:datadistribution} Gaussian Fitting Details~\cite{shi2021predicting}  of Distributions of Adhesive  Free Energies for Data-PS1, Data-PS2, Data-PS3 and Data-PS4}
\end{table}
\end{center}

\subsection*{Transfer Learning Architecture} 

\begin{figure*}[h]
	\begin{center}
	\includegraphics[width=\widefigurewidth]{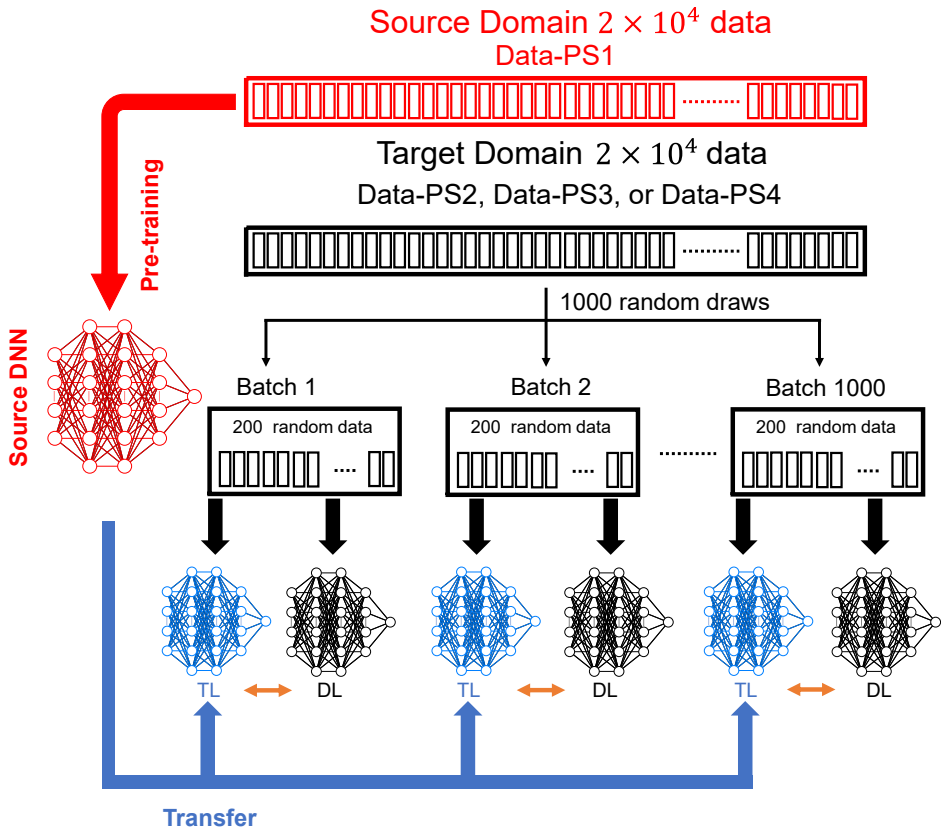}
	\end{center}
	\caption{A schematic of the procedure for testing the performance of transfer learning from source domain (Data-PS1) to 
	 target domain (Data-PS2, Data-PS3, or Data-PS4). A total of $2\times 10 ^4$ polymer sequences and the corresponding $\Delta F$ with PS1 are 
	used as the source data. We train a fully connected deep neural network whose architecture is (20,64,64,32,1), using all the $2\times 10 ^4$ source data points and save its weights. 
	When training the DNN for target data, as a transfer learning framework, 
	we fine-tune a subset of the weights in the pretrained source DNN using 200 data points (TL) and compare with the learning from randomly 
	initialized  DNN in a direct learning (DL) way. 
	}
	\label{fig:TL_model}
\end{figure*}

In this work, a deep neural network (DNN) architecture~\cite{ma2020transfer,briceno2022pemnet} with one input layer, three hidden layers, and one output layer was used to quantify the relationship between the polymer sequence information and polymer--surface adhesive free energy, $\Delta F$. The input was one hot encoding of the polymer sequence. The output was the adhesive free energy. The DNN architecture is shown in Figure ~\ref{fig:TL_model}. 

First, we trained a source DNN with the source data set. We used the Data-PS1, $2\times 10^4$ data points of polymer sequences and their $\Delta F$ with PS1 as the source data,  as the ML model applied to this dataset achieved the highest accuracy among the four original datasets.~\cite{shi2021predicting} Then we randomly separated the $2\times 10^4$ data points into $1.6\times 10^4$ as the training set and $4\times 10^3$ as the validation set. 4:1 ratio is a commonly used ratio in machine learning.~\cite{bishop2006PRML, murphy2012machine, zhang2021dive} The training set is the set of data that was used to train and make the model learn the hidden features/patterns in the data. In each epoch, the same training data was fed to the neural network architecture repeatedly, and the model continued to learn the features of the data. The validation set is a set of data that was used to validate our model performance during training. This validation process provided information that helped tune the model’s hyperparameters and configurations. A test set is not required for this initial task as we are seeking a base line trained on PS1 to extend to the other datasets. Without the need to leave data points for a test set, we were able to have more data points for training and validation. The hyperparameters of the DNN are optimized on the source task PS1 by promoting the accuracy and robustness of the DNN. Utilizing an $n$-tuple description for the hidden layers of a fully connected DNN, our network was represented by (20,64,64,32,1). The learning rate, which serves as the step size for updating the DNN parameters, was set to 0.00002 to make the learning process stable. LeakyReLU~\cite{xu2015empirical} with a negative slope of 0.1 was used as the activation function, and Adam algorithm~\cite{kingma2014adam} was used to optimize weights. The number of learning epochs was set to $10^4$, and the training process can be early ended by a converging check function was applied on the validation data to terminate the training process, if appropriate. We trained a source DNN using the training set of source domain and selected the epoch with the highest accuracy on the validation set as the base DNN for subsequent TL task, which were referred to as the pre-trained source DNN (depicted as the red DNN in Figure~\ref{fig:TL_model}). An open-source machine learning framework, Pytorch~\cite{pytorch}, was used to implement the DNN. All the parameters are stored on Github as described in the Code Availability section.

Next, we turned to the target data set and applied the DNN with the same hyperparameters. The small target data set was composed of 200 data points which were randomly drawn from existing data on the new domain (Data-PS2, Data-PS3, or Data-PS4). The data set was then divided into training, validation, and test sets in the ratio of 72:18:10, to be consistent with previous transfer learning studies.~\cite{ma2020transfer} 144 training data points were used for training the model, and 36 validation data points were used to determine when the training should be stopped and to avoid overfitting. The validation data set was used to select the training epoch. Since the validation data set was involved in the training process, the model's performance is toward it. Therefore, we additionally tested our model on the untouched test data set to provide unbiased final model performance metrics. Our use  of this protocol enabled us to address the core question: ``How well does the model perform on the small data set of Data-PS2, Data-PS3, or Data-PS4 without bias?'' To illustrate the power of transfer learning, with the same 200 data points and the same separation for training, validation, and test sets, we performed direct learning (DL) (black DNN in Figure ~\ref{fig:TL_model}) and transfer learning (TL) (blue DNN in Figure ~\ref{fig:TL_model}). For direct learning, we trained the DNN model from randomly initialized weights. For transfer learning (blue DNN in Figure ~\ref{fig:TL_model}), we alternatively fine-tuneed the weights of all layers in the pre-trained DNN from the source task. There are three reasons that we choose to fine-tune the weights of all layers. First, we sought to build an end-to-end model which is more friendly to other users who are not familiar with deep learning. In an end-to-end model, users only need to focus on the input and output and do not need to worry about how to modify the inside architecture of the model. We want to show that starting from a pre-trained DNN without fixing the weights can get improvements. Second, we tested other fine-tuning formats, like fixing the weights of the first $n$ layers and fine-tuning the weights of the remaining $m$ layers.\cite{ma2020transfer} We found that those formats do not provide competitive improvements and sometimes behaved worse than when fine-tuning all layers. Third, when the size of the training set increases, fixing the weights of some layers might lead to underfitting. Fine-tuning the weights of all layers is more robust to the size of the training data. 

The comparison between performances of DL and TL scenarios was evaluated by comparing their respective coefficients of determination ($R^2$ values) on the same test sets (20 data points). 
\begin{equation}
    R^2 = 1-\frac{\sum(y_i -\hat{y_i})^2}{\sum(y_i - \Bar{y_i})^2}
\end{equation}
The maximum performance score of $R^2 = 1.0$ occurs when every prediction is correct ($y_i  \equiv \hat{y_i}$). Note that $R^2$ can be negative because the model can be arbitrarily poor. The choice of $R^2$ as an evaluation metric was reasonable, as $R^2$ can provide a natural baseline for judging the performance of models.~\cite{ma2020transfer} For each small data set, we obtained two coefficients: $R^2_{\rm DL}$, which shows the performance of DL, and $R^2_{\rm TL}$ which characterizes the performance of TL. The small data set resulted in highly variable accuracies in models due to the random data drawing. Therefore, we did not limit testing to a single small target data case and randomly drew sample data from the target space 1000 times for both DL and TL scenarios, subsequently obtaining 1000 pairs of $R^2_{\rm DL}$ and $R^2_{\rm TL}$. This enabled us to gain a statistically robust understanding of the behavior of TL, mitigating the effects of outlier data sets on training.

\section*{Results and Discussion}
A summary of performance (captured via $R^2$ values) of direct (DL) and transfer (TL) learning for 1000 small target data sets from Data-PS2, Data-PS3, and Data-PS4 is given in Table~\ref{tab:r2results}. We step through specific cases below.

\begin{center}
\begin{table}
\begin{tabular}{|c|c|c|c|}
\hline
Dataset & $R^2_{\rm DL}$  & $R^2_{\rm TL}$\\
\hline
Data-PS2 & $-0.0089 \pm 0.1956$ & $0.8303 \pm 0.0747$ \\
\hline
Data-PS3 & $0.6338 \pm 0.2079$ & $0.7998 \pm 0.1173$  \\
\hline
Data-PS4 & $0.4502 \pm 0.1849$ & $0.6578 \pm 0.1341$  \\
\hline
\end{tabular}
\caption{\label{tab:r2results} $R^2$ characteristics from DL and TL for 1000 small target data sets for datasets Data-PS2, Data-PS3 and Data-PS4; transfer learning proceeds using a neural network trained on Data-PS1 applied to data on the target surface.}
\end{table}
\end{center}
%\noindent
\subsection*{Knowledge Transfer from Data-PS1 to Data-PS2}

\begin{figure}[h]
	\begin{center}
	\includegraphics[width=.45\textwidth]{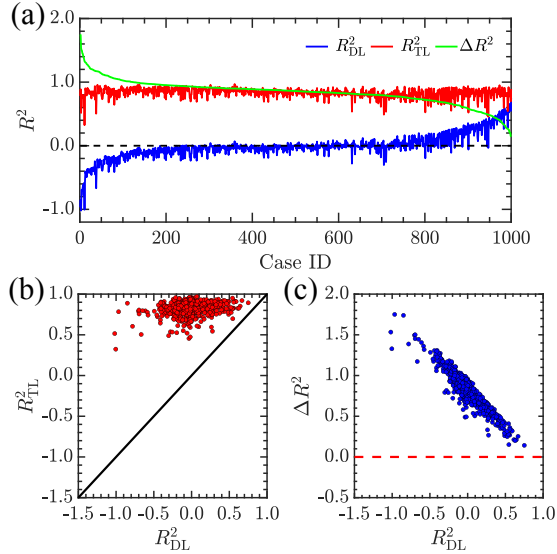}
	\end{center}
	\caption{Transfer learning applied adhesive free energies of sequence defined polymers using a DNN fit to Data-PS1 adapted to Data-PS2 using a small dataset. (a) 1000 pairs of $R^2_{\rm DL}$ for direct learning (blue line) and $R^2_{\rm TL}$ for transfer learning (red line). 
	The improvement from transfer learning ($\Delta R^2 = R^2_{\rm TL} - R^2_{\rm DL}$ is shown in green line.
	The Case ID numbers on the $x$ axis are sorted by the value $\Delta R^2$ in descending order.
	(b) $R^2_{\rm TL}$ plotted against $R^2_{\rm DL}$.
	(c) Improvement $\Delta R^2$ as a function of $R^2_{\rm DL}$.
	}
	\label{fig:TL_PS2}
\end{figure}

We first investigate the application of TL in transferring knowledge from Data-PS1 over to Data-PS2, which links data acquired on from one regular patterned surface to another regular patterned surface. PS1 and PS2 have the same overall composition (red:green = 200:200), and similar patterning when accounting for periodic boundaries, though PS2 uses smaller squares of uniform chemistry rather than two large stripes. The results of 1000 trials are plotted as pairs $R^2_{\rm DL}$ and $R^2_{\rm TL}$ for comparison in Figure~\ref{fig:TL_PS2}.
As shown in Figure~\ref{fig:TL_PS2}(a),  624 DL $R^2$ values are negative, implying poor model performance for those cases. All  $R^2$ values for TL cases are positively correlated, and many are close to one, meaning that the models' performances are excellent in those cases. Collectively, the average $R^2$ on the 1000 test sets through DL is $-0.0089 \pm 0.1956$, while the same metric for TL is $0.8303 \pm 0.0747$. Therefore, TL both improved the mean value of $R^2$ and decreased its standard deviation (SD). The diminishing SD shows that TL is less sensitive to the random selection of the small dataset than the DL, which can be ascribed to the weights of the pre-trained DNN being close to the optimized weights of the target DNN. From the dashed line depicting $\Delta R^2$ in Figure~\ref{fig:TL_PS2}(c), where all the $\Delta R^2$ are greater than zero, and Figure~\ref{fig:TL_PS2}(b), 
where all points ($R^2_{\rm TL}$ vs. $R^2_{\rm DL}$) are above the line $y=x$, it can be inferred that in all the 1000 target cases, TL improved the accuracy of the model prediction.
In Figure~\ref{fig:TL_PS2}(c), a strong negative linear relationship between the improvement in $\Delta R^2$ and the model accuracy from DL demonstrated that TL contributed improved knowledge in situations where DL yielded low accuracy. Since TL transfers a pretrained network rather than initializing weights randomly, the additional small data set acts to refine the weights rather than generate them wholesale --- hence, even when DL yields low accuracy the performance of TL remains stable, as is strongly evident in the the improvement of $\Delta R^2$ for these surfaces. At the same time, when DL already achieved very high model accuracy on the target tasks, the transfer knowledge from the source task offered only a slight improvement. For these cases, the randomly initialized weights of the DNN for DL happened to be close to the optimized weights. 

\subsection*{Knowledge Transfer from Data-PS1 to Data-PS3} 

\begin{figure}[h]
	\begin{center}
	\includegraphics[width=.45\textwidth]{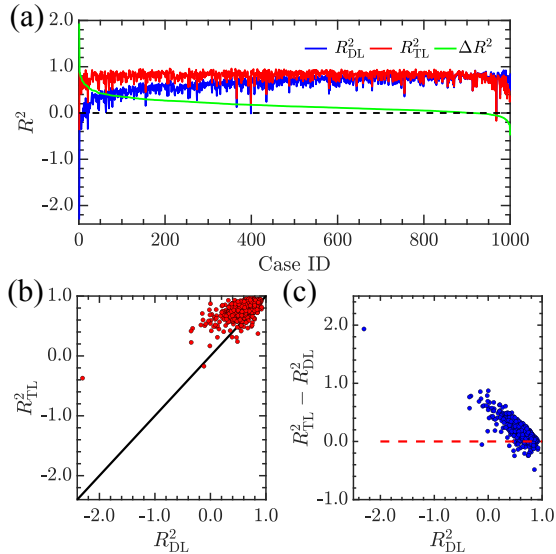}
	\end{center}
	\caption{Transfer learning applied adhesive free energies of sequence defined polymers using a DNN fit to Data-PS1 adapted to Data-PS3 using a small dataset. (a) $R^2$ values for direct learning ($R^2_{\rm DL}$, blue line), transfer learning ($R^2_{\rm TL}$, 
	red line) and improvement from transfer learning ($\Delta R^2 = R^2_{\rm TL} - R^2_{\rm DL}$, green line) of the 1000 target cases. 
	 Case ID numbers on the $x$ axis are sorted by the value $\Delta R^2$ in descending order.
	(b) $R^2_{\rm TL}$ plotted against $R^2_{\rm DL}$.
	(c) Improvement $\Delta R^2$ as a function of $R^2_{\rm DL}$.
	}
	\label{fig:TL_PS3}
\end{figure}

Next, we investigated the application of TL from Data-PS1 to Data-PS3. While PS1 is very regular, PS3 is a fully randomized surface with composition (red:green = 184:216), generated using a random probability $P(\text{red}) = P(\text{green}) = 0.5$ for the beads in the square lattice. The average $R^2$ on 1000 test sets modeled using DL was $0.6338 \pm 0.2079$, while the same metric from TL was improved to $0.7998 \pm 0.1173$. We note that DL's performance for Data-PS3 was better than that for Data-PS2, attributable to the reason that the standard variation ($\sigma = 0.77 \kT$) of the whole $2\times10^4$ Data-PS3's $\Delta F$ is smaller than that of Data-PS2 ($\sigma = 1.55 \kT$).\cite{shi2021predicting} We conclude that the improvement from TL is less robust on this dataset than Data-PS2, likely because of the tighter distribution for adhesive energies (see Ref.~\onlinecite{shi2021predicting} for context). Still there remains a marked improvement. Another significant reason for the differences in this case is that the source and data sets is the more dissimilar adhesion properties related to the randomization of the surface pattern. Still, DL has 15 cases where $R^2_{\rm DL}$ is not greater than zero, while TL only has 2 cases where  $R^2_{\rm TL}$ is not greater than  zero. The green line in Figure ~\ref{fig:TL_PS3}(a) and data in Figure ~\ref{fig:TL_PS3}(b) show that in most examined cases (910 out of 1000), TL gives positive improvement. In Figure ~\ref{fig:TL_PS3}(c), there is a generally negative linear relationship between $\Delta R^2$ and model accuracy from DL, though the linear relationship is not as strong as the prior dataset in Figure ~\ref{fig:TL_PS2}(c). Thus, we infer that when the target tasks have very high model accuracy from DL already, the transfer of knowledge from the source task does not always help further improve the model accuracy. 

\subsection*{Knowledge Transfer from Data-PS1 to Data-PS4} 

\begin{figure}[h]
	\begin{center}
	\includegraphics[width=.45\textwidth]{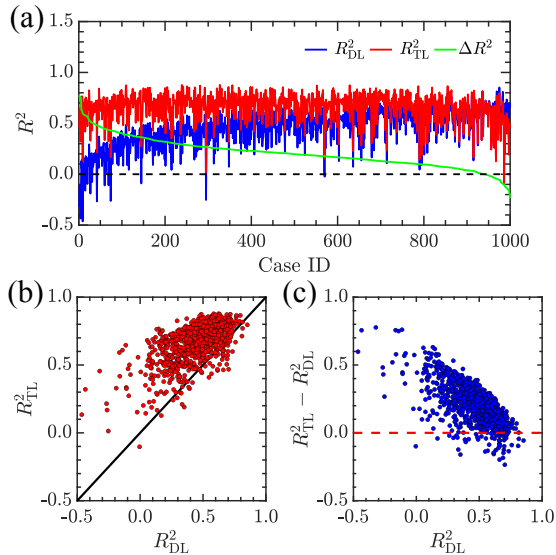}
	\end{center}
	\caption{Transfer learning applied adhesive free energies of sequence defined polymers using a DNN fit to Data-PS1 adapted to Data-PS4 using a small dataset. (a) $R^2$ values for direct learning ($R^2_{\rm DL}$, blue line), transfer learning ($R^2_{\rm TL}$, 
	red line) and improvement from transfer learning ($\Delta R^2 = R^2_{\rm TL} - R^2_{\rm DL}$, green line) of the 1000 target cases. 
	 Case ID numbers on the $x$ axis are sorted by the value $\Delta R^2$ in descending order.
	(b) $R^2_{\rm TL}$ plotted against $R^2_{\rm DL}$.
	(c) Improvement $\Delta R^2$ as a function of $R^2_{\rm DL}$.
	}
	\label{fig:TL_PS4}
\end{figure}

Finally we test the application of TL from Data-PS1 to Data-PS4; the surface  PS4 is a randomized version of PS2 whose composition (red:green = 206:194) differs slightly from the 1:1 composition of PS2.\cite{shi2021predicting} The average $R^2$ on 1000 test sets through DL is $0.4502 \pm 0.1849$, while the same metric from TL is improved to $0.6578 \pm 0.1341$.
DL's performance for Data-PS4 was better than that for Data-PS2, attributable to the relative tightness of the free energy distribution of Data-PS4 ($\sigma = 0.31 \kT$) compared to Data-PS2 ($\sigma = 1.55 \kT$).~\cite{shi2021predicting} The improvement is not as evident as in the first case (Data-PS1 to Data-PS2), but is overall much better than the performance using TL on the completely randomized PS3 surface. The green line in Figure~\ref{fig:TL_PS4}(a) and Figure ~\ref{fig:TL_PS4}(b) show most cases (939 out of 1000) are positively impacted by TL.  In Figure ~\ref{fig:TL_PS4}(c), the negatively correlated relationship between the improvement in $\Delta R^2$ and the model accuracy from DL appeared weaker than the previous two cases in Figure ~\ref{fig:TL_PS2}(c) and Figure ~\ref{fig:TL_PS3}(c), though we note that a single testing point with good TL and poor DL performance skews the plots visually.

\subsection*{Feature Importance Analysis}

The structure of the one-hot encoding of our sequence-defined polymers permitted the interrogation of the feature importance of various sites on the polymer backbone. We utilize the entire data set of Data-PS1, Data-PS2, Data-PS3, and Data-PS4. The details of the training process were identical to those stated in the Methods section. Permutation feature importance, which is defined to be the decrease in predictive accuracy ($\Delta R^2$) when a single feature value is randomly shuffled, was used to evaluate descriptor importance.~\cite{altmann2010permutation,ma2020transfer} The feature importance for the feature $i$ is computed by
\begin{equation*}
    {\rm FI}_{i} = \Delta R^2 =R^2- R^2_{i}\;,
\end{equation*}
where $R^2$ is the predictive accuracy without randomly shuffling and $R^2_{i}$ is the predictive accuracy after randomly shuffling the $i^{\text{th}}$ dimensional feature. We used the permutation feature importance implementation in the Python package ELI5~\cite{ELI5} to perform this analysis. Since our input is the one-hot encoding of the polymer sequence, a 20-dimensional vector, we alternatively shuffled the feature value of each dimension and calculated the descriptor importance for every input variable. 

\begin{figure}
	\begin{center}
	\includegraphics[width=0.65\textwidth]{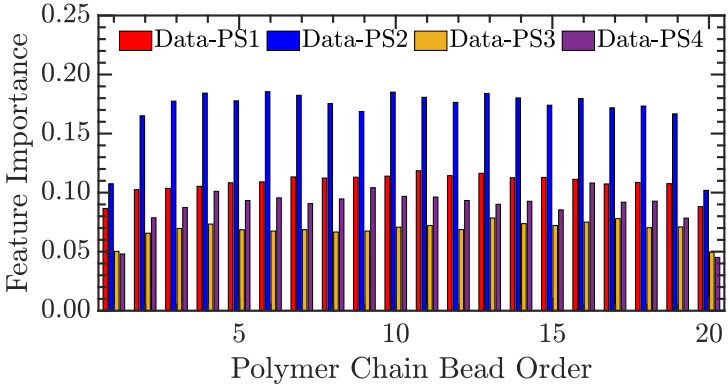}
	\end{center}
	\caption{Permutation feature importance of 20-dimensional input vector representing for bead in the sequence-defined polymer for four data set, Data-PS1 (red), Data-PS2 (blue), Data-PS3 (orange) and Data-PS4 (purple).
	Essentially, only the endpoints are significantly different with the 18 interior beads having rough similar importance to one another. 
	}
	\label{fig:FI}
\end{figure}

The results of the feature importance analysis are shown in Figure~\ref{fig:FI}. Even though the absolute value of feature importance is different for each patterned surface, some common features exist for all four patterned surfaces. The head (first) and the tail (twentieth) beads had relatively lower values of feature importance, and the other eighteen beads' feature importance were almost the same within the individual surface dataset. Statt et al.,~\cite{statt2020model} also found that the ends of an intrinsically disordered protein (IDP) have a distinct effect on the phase behavior (critical temperature) compared with mutations in the middle of the chain, though the ends are seen there to have a more pronounced effect on the proteins' phase behavior. The common features we found among Data-PS1, Data-PS2, Data-PS3, and Data-PS4, represent the shareable knowledge from TL and can explain the successful application of TL in these cases. Pre-trained models were able to obtain these features before fine-tuning with the small datasets.

\subsection*{Size Effects for TL Improvements}

From the above investigations, we illustrated that transfer learning can improve the accuracy of the DNN models trained on a small target dataset (200 data points).
It is of interest to see how this scales with the amount of the available data, thus we also explored the effects of the size of the target data set on the improvements from TL, 
increasing the ``small'' data set to values between 200 and 4000 points. As previously, other training settings were kept the same as for $N=200$ datasets.

\begin{figure}
	\begin{center}
	\includegraphics[width=0.40\textwidth]{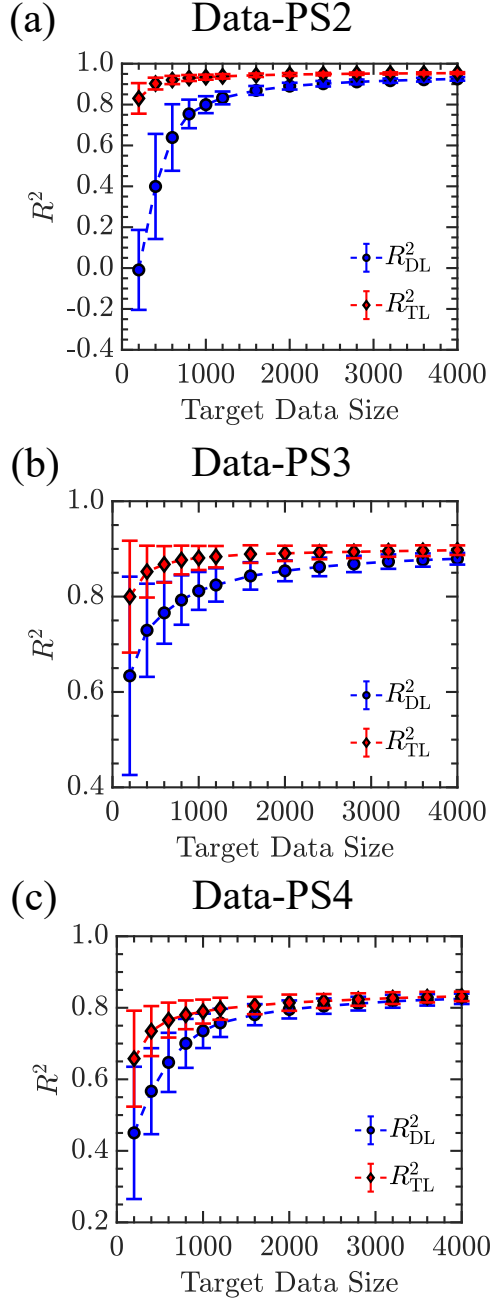}
	\end{center}
	\caption{(a) $R^2_{\rm DL}$ (blue), $R^2_{\rm TL}$ (red)  as a function of the size of the target dataset.
	Data-PS2 was used for this comparison.
	The error bars reflect the SD of the $R^2$ score from 1000 random draws. 
	As the size of the target data set increases, $R^2_{\rm DL}$ and $R^2_{\rm TL}$ both increase and their SD values decrease. $R^2_{\rm DL}$ 
	increases more dramatically relative to $R^2_{\rm TL}$. Though the improvement $\Delta R^2$ decreases with increasing size of the data set, $R^2_{\rm TL}$ is always larger than $R^2_{\rm DL}$.
	Similar improvements of DL and TL were seen in Data-PS3 (b) and Data-PS4 (c), though the behavior of TL saturated at lower accuracy in each relative to Data-PS2.
	}
	\label{fig:size_all}
\end{figure}

Figure~\ref{fig:size_all}(a) illustrates the prediction performance of DL and TL for Data-PS2 as a function of the size of the target data set, as quantified by the $R^2$ score. The mean value increases and the SD decrease for the $ R^2$ score in both DL and TL as the size of the target data set increases. These values are seen to change more rapidly in DL than TL, which is perhaps to be expected, as more data should drastically improve the behavior of DL given the initial datasets are so small; the pretrained TL model is able to approach an optimal fit more easily. We note $R^2_{\rm TL}$ quickly converged with increasing size. After the target data size exceeded 800, accuracy (measured by $R^2_{\rm TL}$) saturated. Thus, the improvement between TL and DL decreased as the target data set size increases. Still, the performance of the TL models was always better on average than DL models. This indicates that the transfer of knowledge from the source task can offer significant improvement when scant target data is available. The DNN model can learn sufficiently directly from the feed data in large target data sets, and TL does not offer significant improvement. In our data sets, $2\times10^4$ independent points are available, and TL's efficacy saturates relative to DL when approximately 20\% of the target data set is use. This suggests a threshold below which TL should always be used. Similar improvements of DL and TL were seen in Data-PS3 and Data-PS4 [see Fig.~\ref{fig:size_all}(b,c)], though the behavior of TL saturated at lower accuracy in each relative to Data-PS2. Nonetheless, improvements on the order of one standard deviation in $R^2$ were seen up to the same 20\% threshold applied to the target data size.

\section*{Conclusion}
In summary, a comprehensive TL study of the polymer--surface interaction between polymers with defined sequences and different surfaces was conducted through pre-training a DNN with a large dataset from the source domain (Data-PS1) and fine-tuning the pre-trained DNN with the small dataset from the target domain (Data-PS2, Data-PS3, or Data-PS4). Knowledge was transferable among the polymer interactions with different patterned surfaces. TL significantly upgraded the performance of the model trained on the small dataset. % ($-0.0089 \pm 0.1956$ to $0.8303 \pm 0.0747$; $0.6338 \pm 0.2079$ to $0.7998 \pm 0.1173$;  $0.4502 \pm 0.1849$ to $0.6578 \pm 0.1341$). 
In addition, our results showed that TL's model is more stable than the DL and less likely to be affected by the random selection of data. The study of permutation feature importance revealed that the four patterned surfaces have some similar features, representing part of the reasons the transferable knowledge can work. We also demonstrate that the increase in target data size can diminish the improvement from TL, ascribable to the fact that DL learns more knowledge from the feed data directly when the size of the target data increases. However, the TL model with the full-fine-tuning architecture always performs better than the DL model, even though the improvement diminishes at large sizes of datasets.

Our work highlights the importance of transfer learning in elevating the performance of an ML model regarding the polymer--surface interaction under insufficient data dilemmas. Usually, tens or hundreds of data points are insufficient to train an accurate ML model. With transfer learning tools, the shareable knowledge from a pre-trained model can help the ML model trained for polymer--surface interaction with a new surface to obtain higher performance. Our test cases are all simulation datasets, and the knowledge is shared among different surfaces in simulation. But the benefit of knowledge sharing is not only limited to simulation data sets. For example, Briceno-Mena \textit{et al.}~\cite{briceno2022pemnet}  have utilized transfer learning techniques to increase the performance of an ML model trained with insufficient experimental data by transferring knowledge from an ML model trained with a large simulation dataset.~\cite{briceno2022pemnet} Similarly, for the prediction and optimization of adhesive energies, transfer learning can be used to maximize our knowledge within a new chemical domain from a smaller amount of simulations or experiments, perhaps allowing purely computational and coarse-grained models to cheaply explore compositional space and predictive models to be refined by directed experimentation. As transfer learning, especially few-shot learning, has achieved dramatic successes in computer vision and language models by building a series of sizeable pre-trained ML models,  such as  YOLO,~\cite{redmon2016you} BERT,\cite{bert}  and GPT-3,~\cite{gpt3} we anticipate knowledge about network structure and problem complexity may be used to guide these algorithms and their applications to materials problems. With more experimental and simulation data about polymer--surface interactions being produced and collected in the future, it is expected to obtain a sizeable pre-trained ML model for polymer surface interaction. 

\section*{Data Availability}
The adhesive free energy datasets that used in this article are available online at \url{https://github.com/shijiale0609/ML_PSI}. Example scripts and information necessary to run the examples contained in this article are posted at \url{https://github.com/shijiale0609/TL_PSI}.

\section*{Author Contributions}
JS, FA, SS, and JKW conceived the study. JS and JKW designed the research plan. YJC provided suggestions and guidance on the transfer learning framework. JS conducted transfer learning model training and analysed the results. JS, FA, YJC, SS, and JKW interpreted the results and wrote the paper.

\section*{Acknowledgement}
JS, and JKW acknowledge the support of MICCoM, the Midwest Center for Computational Materials, as part of the Computational Materials Sciences Program funded by the U.S.  Department of Energy, Office of Science, Basic Energy Sciences, Materials Sciences  and Engineering Division, for the development of algorithms and codes used within this work. JS, and JKW acknowledge computational resources at the Notre Dame Center for Research Computing (CRC). YJC acknowledges support from the US National Science Foundation Faculty Early Career Development Program (CAREER) through award CBET-2143346.

\bibliography{ref}

\providecommand{\latin}[1]{#1}
\makeatletter
\providecommand{\doi}
  {\begingroup\let\do\@makeother\dospecials
  \catcode`\{=1 \catcode`\}=2 \doi@aux}
\providecommand{\doi@aux}[1]{\endgroup\texttt{#1}}
\makeatother
\providecommand*\mcitethebibliography{\thebibliography}
\csname @ifundefined\endcsname{endmcitethebibliography}
  {\let\endmcitethebibliography\endthebibliography}{}
\begin{mcitethebibliography}{77}
\providecommand*\natexlab[1]{#1}
\providecommand*\mciteSetBstSublistMode[1]{}
\providecommand*\mciteSetBstMaxWidthForm[2]{}
\providecommand*\mciteBstWouldAddEndPuncttrue
  {\def\EndOfBibitem{\unskip.}}
\providecommand*\mciteBstWouldAddEndPunctfalse
  {\let\EndOfBibitem\relax}
\providecommand*\mciteSetBstMidEndSepPunct[3]{}
\providecommand*\mciteSetBstSublistLabelBeginEnd[3]{}
\providecommand*\EndOfBibitem{}
\mciteSetBstSublistMode{f}
\mciteSetBstMaxWidthForm{subitem}{(\alph{mcitesubitemcount})}
\mciteSetBstSublistLabelBeginEnd
  {\mcitemaxwidthsubitemform\space}
  {\relax}
  {\relax}

\bibitem[Chakraborty and Golumbfskie(2001)Chakraborty, and
  Golumbfskie]{chakraborty2001polymer}
Chakraborty,~A.~K.; Golumbfskie,~A.~J. Polymer Adsorption--Driven Self-Assembly
  of Nanostructures. \emph{Annu. Rev. Phys. Chem.} \textbf{2001}, \emph{52},
  537--573, PMID: 11326074\relax
\mciteBstWouldAddEndPuncttrue
\mciteSetBstMidEndSepPunct{\mcitedefaultmidpunct}
{\mcitedefaultendpunct}{\mcitedefaultseppunct}\relax
\EndOfBibitem
\bibitem[Li \latin{et~al.}(2019)Li, Lin, Tang, Duncan, and Ke]{li2019advanced}
Li,~L.; Lin,~Q.; Tang,~M.; Duncan,~A. J.~E.; Ke,~C. Advanced Polymer Designs
  for Direct-Ink-Write 3D Printing. \emph{Chemistry – A European Journal}
  \textbf{2019}, \emph{25}, 10768--10781\relax
\mciteBstWouldAddEndPuncttrue
\mciteSetBstMidEndSepPunct{\mcitedefaultmidpunct}
{\mcitedefaultendpunct}{\mcitedefaultseppunct}\relax
\EndOfBibitem
\bibitem[Kim \latin{et~al.}(2003)Kim, Solak, Stoykovich, Ferrier, De~Pablo, and
  Nealey]{kim2003epitaxial}
Kim,~S.~O.; Solak,~H.~H.; Stoykovich,~M.~P.; Ferrier,~N.~J.; De~Pablo,~J.~J.;
  Nealey,~P.~F. Epitaxial self-assembly of block copolymers on lithographically
  defined nanopatterned substrates. \emph{Nature} \textbf{2003}, \emph{424},
  411--414\relax
\mciteBstWouldAddEndPuncttrue
\mciteSetBstMidEndSepPunct{\mcitedefaultmidpunct}
{\mcitedefaultendpunct}{\mcitedefaultseppunct}\relax
\EndOfBibitem
\bibitem[Ikawa \latin{et~al.}(2012)Ikawa, Yamada, Matsui, Minemawari, Tsutsumi,
  Horii, Chikamatsu, Azumi, Kumai, and Hasegawa]{ikawa2012simple}
Ikawa,~M.; Yamada,~T.; Matsui,~H.; Minemawari,~H.; Tsutsumi,~J.; Horii,~Y.;
  Chikamatsu,~M.; Azumi,~R.; Kumai,~R.; Hasegawa,~T. Simple push coating of
  polymer thin-film transistors. \emph{Nature Communications} \textbf{2012},
  \emph{3}, 1--8\relax
\mciteBstWouldAddEndPuncttrue
\mciteSetBstMidEndSepPunct{\mcitedefaultmidpunct}
{\mcitedefaultendpunct}{\mcitedefaultseppunct}\relax
\EndOfBibitem
\bibitem[Annabi \latin{et~al.}(2014)Annabi, Tamayol, Shin, Ghaemmaghami,
  Peppas, and Khademhosseini]{Annabi2014}
Annabi,~N.; Tamayol,~A.; Shin,~S.~R.; Ghaemmaghami,~A.~M.; Peppas,~N.~A.;
  Khademhosseini,~A. Surgical materials: Current challenges and nano-enabled
  solutions. \emph{Nano Today} \textbf{2014}, \emph{9}, 574--589\relax
\mciteBstWouldAddEndPuncttrue
\mciteSetBstMidEndSepPunct{\mcitedefaultmidpunct}
{\mcitedefaultendpunct}{\mcitedefaultseppunct}\relax
\EndOfBibitem
\bibitem[Xiu \latin{et~al.}(2020)Xiu, Dick, Ju, Mirzaie, Abdi, Cocklin, Zhan,
  and Liu]{xiu2020Inhibitors}
Xiu,~S.; Dick,~A.; Ju,~H.; Mirzaie,~S.; Abdi,~F.; Cocklin,~S.; Zhan,~P.;
  Liu,~X. Inhibitors of SARS-CoV-2 Entry: Current and Future Opportunities.
  \emph{J. Med. Chem} \textbf{2020}, \emph{63}, 12256--12274\relax
\mciteBstWouldAddEndPuncttrue
\mciteSetBstMidEndSepPunct{\mcitedefaultmidpunct}
{\mcitedefaultendpunct}{\mcitedefaultseppunct}\relax
\EndOfBibitem
\bibitem[Wong and Damania(2021)Wong, and Damania]{wong2021sars}
Wong,~J.~P.; Damania,~B. SARS-CoV-2 dependence on host pathways. \emph{Science}
  \textbf{2021}, \emph{371}, 884--885\relax
\mciteBstWouldAddEndPuncttrue
\mciteSetBstMidEndSepPunct{\mcitedefaultmidpunct}
{\mcitedefaultendpunct}{\mcitedefaultseppunct}\relax
\EndOfBibitem
\bibitem[Callaway(2020)]{callaway2020making}
Callaway,~E. Making sense of coronavirus mutations. \emph{Nature}
  \textbf{2020}, 174--177\relax
\mciteBstWouldAddEndPuncttrue
\mciteSetBstMidEndSepPunct{\mcitedefaultmidpunct}
{\mcitedefaultendpunct}{\mcitedefaultseppunct}\relax
\EndOfBibitem
\bibitem[Plante \latin{et~al.}(2021)Plante, Liu, Liu, Xia, Johnson, Lokugamage,
  Zhang, Muruato, Zou, Fontes-Garfias, Mirchandani, Scharton, Bilello, Ku, An,
  Kalveram, Freiberg, Menachery, Xie, Plante, Weaver, and Shi]{plante2021spike}
Plante,~J.~A.; Liu,~Y.; Liu,~J.; Xia,~H.; Johnson,~B.~A.; Lokugamage,~K.~G.;
  Zhang,~X.; Muruato,~A.~E.; Zou,~J.; Fontes-Garfias,~C.~R.; Mirchandani,~D.;
  Scharton,~D.; Bilello,~J.~P.; Ku,~Z.; An,~Z.; Kalveram,~B.; Freiberg,~A.~N.;
  Menachery,~V.~D.; Xie,~X.; Plante,~K.~S.; Weaver,~S.~C.; Shi,~P.-Y. {Spike
  mutation D614G alters SARS-CoV-2 fitness}. \emph{Nature} \textbf{2021},
  \emph{592}, 116--121\relax
\mciteBstWouldAddEndPuncttrue
\mciteSetBstMidEndSepPunct{\mcitedefaultmidpunct}
{\mcitedefaultendpunct}{\mcitedefaultseppunct}\relax
\EndOfBibitem
\bibitem[Hie \latin{et~al.}(2021)Hie, Zhong, Berger, and
  Bryson]{hie2021learning}
Hie,~B.; Zhong,~E.~D.; Berger,~B.; Bryson,~B. Learning the language of viral
  evolution and escape. \emph{Science} \textbf{2021}, \emph{371},
  284--288\relax
\mciteBstWouldAddEndPuncttrue
\mciteSetBstMidEndSepPunct{\mcitedefaultmidpunct}
{\mcitedefaultendpunct}{\mcitedefaultseppunct}\relax
\EndOfBibitem
\bibitem[Ozboyaci \latin{et~al.}(2016)Ozboyaci, Kokh, Corni, and
  Wade]{ozboyaci2016modeling}
Ozboyaci,~M.; Kokh,~D.~B.; Corni,~S.; Wade,~R.~C. Modeling and simulation of
  protein--surface interactions: achievements and challenges. \emph{Quarterly
  Reviews of Biophysics} \textbf{2016}, \emph{49}\relax
\mciteBstWouldAddEndPuncttrue
\mciteSetBstMidEndSepPunct{\mcitedefaultmidpunct}
{\mcitedefaultendpunct}{\mcitedefaultseppunct}\relax
\EndOfBibitem
\bibitem[Chakraborty and Bratko(1998)Chakraborty, and
  Bratko]{chakraborty1998simple}
Chakraborty,~A.~K.; Bratko,~D. A simple theory and Monte Carlo simulations for
  recognition between random heteropolymers and disordered surfaces. \emph{J.
  Chem. Phys.} \textbf{1998}, \emph{108}, 1676--1682\relax
\mciteBstWouldAddEndPuncttrue
\mciteSetBstMidEndSepPunct{\mcitedefaultmidpunct}
{\mcitedefaultendpunct}{\mcitedefaultseppunct}\relax
\EndOfBibitem
\bibitem[Muthukumar(1995)]{muthukumar1995pattern}
Muthukumar,~M. Pattern recognition by polyelectrolytes. \emph{J. Chem. Phys.}
  \textbf{1995}, \emph{103}, 4723--4731\relax
\mciteBstWouldAddEndPuncttrue
\mciteSetBstMidEndSepPunct{\mcitedefaultmidpunct}
{\mcitedefaultendpunct}{\mcitedefaultseppunct}\relax
\EndOfBibitem
\bibitem[Chakraborty(2001)]{chakraborty2001disordered}
Chakraborty,~A.~K. Disordered heteropolymers: models for biomimetic polymers
  and polymers with frustrating quenched disorder. \emph{Phys. Rep.}
  \textbf{2001}, \emph{342}, 1--61\relax
\mciteBstWouldAddEndPuncttrue
\mciteSetBstMidEndSepPunct{\mcitedefaultmidpunct}
{\mcitedefaultendpunct}{\mcitedefaultseppunct}\relax
\EndOfBibitem
\bibitem[Muthukumar(1998)]{Muthukumar1998pattern}
Muthukumar,~M. Pattern recognition in self-assembly. \emph{Curr. Opin. Colloid
  Interface Sci.} \textbf{1998}, \emph{3}, 48--54\relax
\mciteBstWouldAddEndPuncttrue
\mciteSetBstMidEndSepPunct{\mcitedefaultmidpunct}
{\mcitedefaultendpunct}{\mcitedefaultseppunct}\relax
\EndOfBibitem
\bibitem[Chauhan \latin{et~al.}(2021)Chauhan, Simpson, and
  Abel]{Chauhan2021crowding}
Chauhan,~G.; Simpson,~M.~L.; Abel,~S.~M. Crowding-induced interactions of ring
  polymers. \emph{Soft Matter} \textbf{2021}, \emph{17}, 16--23\relax
\mciteBstWouldAddEndPuncttrue
\mciteSetBstMidEndSepPunct{\mcitedefaultmidpunct}
{\mcitedefaultendpunct}{\mcitedefaultseppunct}\relax
\EndOfBibitem
\bibitem[Kriksin \latin{et~al.}(2005)Kriksin, Khalatur, and
  Khokhlov]{kriksin2005adsorption}
Kriksin,~Y.~A.; Khalatur,~P.~G.; Khokhlov,~A.~R. Adsorption of multiblock
  copolymers onto a chemically heterogeneous surface: A model of pattern
  recognition. \emph{J. Chem. Phys.} \textbf{2005}, \emph{122}, 114703\relax
\mciteBstWouldAddEndPuncttrue
\mciteSetBstMidEndSepPunct{\mcitedefaultmidpunct}
{\mcitedefaultendpunct}{\mcitedefaultseppunct}\relax
\EndOfBibitem
\bibitem[Tereshkin \latin{et~al.}(2022)Tereshkin, Tereshkina, and
  Krupyanskii]{tereshkin2022predicting}
Tereshkin,~E.~V.; Tereshkina,~K.~B.; Krupyanskii,~Y.~F. Predicting Binding Free
  Energies for DPS Protein-DNA Complexes and Crystals Using Molecular Dynamics.
  \emph{Supercomputing Frontiers and Innovations} \textbf{2022}, \emph{9},
  33--45\relax
\mciteBstWouldAddEndPuncttrue
\mciteSetBstMidEndSepPunct{\mcitedefaultmidpunct}
{\mcitedefaultendpunct}{\mcitedefaultseppunct}\relax
\EndOfBibitem
\bibitem[Shi \latin{et~al.}(2022)Shi, Quevillon, Amorim~Valença, and
  Whitmer]{shi2021predicting}
Shi,~J.; Quevillon,~M.~J.; Amorim~Valença,~P.~H.; Whitmer,~J.~K. Predicting
  Adhesive Free Energies of Polymer–Surface Interactions with Machine
  Learning. \emph{ACS Applied Materials \& Interfaces} \textbf{2022},
  \emph{14}, 37161--37169\relax
\mciteBstWouldAddEndPuncttrue
\mciteSetBstMidEndSepPunct{\mcitedefaultmidpunct}
{\mcitedefaultendpunct}{\mcitedefaultseppunct}\relax
\EndOfBibitem
\bibitem[Bishop(2006)]{bishop2006PRML}
Bishop,~C.~M. \emph{Pattern Recognition and Machine Learning (Information
  Science and Statistics)}; Springer-Verlag: Berlin, Heidelberg, 2006\relax
\mciteBstWouldAddEndPuncttrue
\mciteSetBstMidEndSepPunct{\mcitedefaultmidpunct}
{\mcitedefaultendpunct}{\mcitedefaultseppunct}\relax
\EndOfBibitem
\bibitem[Murphy(2012)]{murphy2012machine}
Murphy,~K.~P. \emph{Machine learning: a probabilistic perspective}; MIT press,
  2012\relax
\mciteBstWouldAddEndPuncttrue
\mciteSetBstMidEndSepPunct{\mcitedefaultmidpunct}
{\mcitedefaultendpunct}{\mcitedefaultseppunct}\relax
\EndOfBibitem
\bibitem[Murphy(2022)]{pml1Book}
Murphy,~K.~P. \emph{Probabilistic Machine Learning: An introduction}; MIT
  Press, 2022\relax
\mciteBstWouldAddEndPuncttrue
\mciteSetBstMidEndSepPunct{\mcitedefaultmidpunct}
{\mcitedefaultendpunct}{\mcitedefaultseppunct}\relax
\EndOfBibitem
\bibitem[Murphy(2023)]{pml2Book}
Murphy,~K.~P. \emph{Probabilistic Machine Learning: Advanced Topics}; MIT
  Press, 2023\relax
\mciteBstWouldAddEndPuncttrue
\mciteSetBstMidEndSepPunct{\mcitedefaultmidpunct}
{\mcitedefaultendpunct}{\mcitedefaultseppunct}\relax
\EndOfBibitem
\bibitem[Artrith \latin{et~al.}(2021)Artrith, Butler, Coudert, Han, Isayev,
  Jain, and Walsh]{artrith2021best}
Artrith,~N.; Butler,~K.~T.; Coudert,~F.-X.; Han,~S.; Isayev,~O.; Jain,~A.;
  Walsh,~A. Best practices in machine learning for chemistry. \emph{Nat. Chem.}
  \textbf{2021}, \emph{13}, 505--508\relax
\mciteBstWouldAddEndPuncttrue
\mciteSetBstMidEndSepPunct{\mcitedefaultmidpunct}
{\mcitedefaultendpunct}{\mcitedefaultseppunct}\relax
\EndOfBibitem
\bibitem[de~Pablo \latin{et~al.}(2019)de~Pablo, Jackson, Webb, Chen, Moore,
  Morgan, Jacobs, Pollock, Schlom, Toberer, Analytis, Dabo, DeLongchamp, Fiete,
  Grason, Hautier, Mo, Rajan, Reed, Rodriguez, Stevanovic, Suntivich, Thornton,
  and Zhao]{de2019new}
de~Pablo,~J.~J.; Jackson,~N.~E.; Webb,~M.~A.; Chen,~L.-Q.; Moore,~J.~E.;
  Morgan,~D.; Jacobs,~R.; Pollock,~T.; Schlom,~D.~G.; Toberer,~E.~S.;
  Analytis,~J.; Dabo,~I.; DeLongchamp,~D.~M.; Fiete,~G.~A.; Grason,~G.~M.;
  Hautier,~G.; Mo,~Y.; Rajan,~K.; Reed,~E.~J.; Rodriguez,~E.; Stevanovic,~V.;
  Suntivich,~J.; Thornton,~K.; Zhao,~J.-C. {New frontiers for the materials
  genome initiative}. \emph{npj Computational Materials} \textbf{2019},
  \emph{5}, 41\relax
\mciteBstWouldAddEndPuncttrue
\mciteSetBstMidEndSepPunct{\mcitedefaultmidpunct}
{\mcitedefaultendpunct}{\mcitedefaultseppunct}\relax
\EndOfBibitem
\bibitem[Gormley and Webb(2021)Gormley, and Webb]{gormley2021machine}
Gormley,~A.~J.; Webb,~M.~A. Machine learning in combinatorial polymer
  chemistry. \emph{Nature Reviews Materials} \textbf{2021}, 1--3\relax
\mciteBstWouldAddEndPuncttrue
\mciteSetBstMidEndSepPunct{\mcitedefaultmidpunct}
{\mcitedefaultendpunct}{\mcitedefaultseppunct}\relax
\EndOfBibitem
\bibitem[Huang \latin{et~al.}(2020)Huang, Fu, Glass, Zitnik, Xiao, and
  Sun]{huang2020deepPurpose}
Huang,~K.; Fu,~T.; Glass,~L.~M.; Zitnik,~M.; Xiao,~C.; Sun,~J. {DeepPurpose: a
  deep learning library for drug–target interaction prediction}.
  \emph{Bioinformatics} \textbf{2020}, \emph{36}, 5545--5547\relax
\mciteBstWouldAddEndPuncttrue
\mciteSetBstMidEndSepPunct{\mcitedefaultmidpunct}
{\mcitedefaultendpunct}{\mcitedefaultseppunct}\relax
\EndOfBibitem
\bibitem[Liang \latin{et~al.}(2022)Liang, Li, Zhou, Sun, Yuan, and
  Zhang]{liang2022machine}
Liang,~Z.; Li,~Z.; Zhou,~S.; Sun,~Y.; Yuan,~J.; Zhang,~C. Machine-learning
  exploration of polymer compatibility. \emph{Cell Reports Physical Science}
  \textbf{2022}, \emph{3}, 100931\relax
\mciteBstWouldAddEndPuncttrue
\mciteSetBstMidEndSepPunct{\mcitedefaultmidpunct}
{\mcitedefaultendpunct}{\mcitedefaultseppunct}\relax
\EndOfBibitem
\bibitem[Briceno-Mena \latin{et~al.}(2022)Briceno-Mena, Romagnoli, and
  Arges]{briceno2022pemnet}
Briceno-Mena,~L.~A.; Romagnoli,~J.~A.; Arges,~C.~G. PemNet: A Transfer
  Learning-Based Modeling Approach of High-Temperature Polymer Electrolyte
  Membrane Electrochemical Systems. \emph{Industrial \& Engineering Chemistry
  Research} \textbf{2022}, \emph{61}, 3350--3357\relax
\mciteBstWouldAddEndPuncttrue
\mciteSetBstMidEndSepPunct{\mcitedefaultmidpunct}
{\mcitedefaultendpunct}{\mcitedefaultseppunct}\relax
\EndOfBibitem
\bibitem[Sattari \latin{et~al.}(2021)Sattari, Xie, and Lin]{sattari2021data}
Sattari,~K.; Xie,~Y.; Lin,~J. Data-driven algorithms for inverse design of
  polymers. \emph{Soft Matter} \textbf{2021}, \emph{17}, 7607--7622\relax
\mciteBstWouldAddEndPuncttrue
\mciteSetBstMidEndSepPunct{\mcitedefaultmidpunct}
{\mcitedefaultendpunct}{\mcitedefaultseppunct}\relax
\EndOfBibitem
\bibitem[Sevgen \latin{et~al.}(2020)Sevgen, Guo, Sidky, Whitmer, and
  de~Pablo]{Sevgen2020combined}
Sevgen,~E.; Guo,~A.~Z.; Sidky,~H.; Whitmer,~J.~K.; de~Pablo,~J.~J. Combined
  Force-Frequency Sampling for Simulation of Systems Having Rugged Free Energy
  Landscapes. \emph{Journal of Chemical Theory and Computation} \textbf{2020},
  \emph{16}, 1448--1455, PMID: 31951703\relax
\mciteBstWouldAddEndPuncttrue
\mciteSetBstMidEndSepPunct{\mcitedefaultmidpunct}
{\mcitedefaultendpunct}{\mcitedefaultseppunct}\relax
\EndOfBibitem
\bibitem[Sidky and Whitmer(2018)Sidky, and Whitmer]{sidky2018learning}
Sidky,~H.; Whitmer,~J.~K. Learning free energy landscapes using artificial
  neural networks. \emph{The Journal of Chemical Physics} \textbf{2018},
  \emph{148}, 104111\relax
\mciteBstWouldAddEndPuncttrue
\mciteSetBstMidEndSepPunct{\mcitedefaultmidpunct}
{\mcitedefaultendpunct}{\mcitedefaultseppunct}\relax
\EndOfBibitem
\bibitem[Jumper \latin{et~al.}(2021)Jumper, Evans, Pritzel, Green, Figurnov,
  Ronneberger, Tunyasuvunakool, Bates, {\v{Z}}{\'{i}}dek, Potapenko, Bridgland,
  Meyer, Kohl, Ballard, Cowie, Romera-Paredes, Nikolov, Jain, Adler, Back,
  Petersen, Reiman, Clancy, Zielinski, Steinegger, Pacholska, Berghammer,
  Bodenstein, Silver, Vinyals, Senior, Kavukcuoglu, Kohli, and
  Hassabis]{alphafold2}
Jumper,~J.; Evans,~R.; Pritzel,~A.; Green,~T.; Figurnov,~M.; Ronneberger,~O.;
  Tunyasuvunakool,~K.; Bates,~R.; {\v{Z}}{\'{i}}dek,~A.; Potapenko,~A.;
  Bridgland,~A.; Meyer,~C.; Kohl,~S. A.~A.; Ballard,~A.~J.; Cowie,~A.;
  Romera-Paredes,~B.; Nikolov,~S.; Jain,~R.; Adler,~J.; Back,~T.; Petersen,~S.;
  Reiman,~D.; Clancy,~E.; Zielinski,~M.; Steinegger,~M.; Pacholska,~M.;
  Berghammer,~T.; Bodenstein,~S.; Silver,~D.; Vinyals,~O.; Senior,~A.~W.;
  Kavukcuoglu,~K.; Kohli,~P.; Hassabis,~D. {Highly accurate protein structure
  prediction with AlphaFold}. \emph{Nature} \textbf{2021}, \emph{596},
  583--589\relax
\mciteBstWouldAddEndPuncttrue
\mciteSetBstMidEndSepPunct{\mcitedefaultmidpunct}
{\mcitedefaultendpunct}{\mcitedefaultseppunct}\relax
\EndOfBibitem
\bibitem[Tunyasuvunakool \latin{et~al.}(2021)Tunyasuvunakool, Adler, Wu, Green,
  Zielinski, {\v{Z}}{\'{i}}dek, Bridgland, Cowie, Meyer, Laydon, Velankar,
  Kleywegt, Bateman, Evans, Pritzel, Figurnov, Ronneberger, Bates, Kohl,
  Potapenko, Ballard, Romera-Paredes, Nikolov, Jain, Clancy, Reiman, Petersen,
  Senior, Kavukcuoglu, Birney, Kohli, Jumper, and
  Hassabis]{tunyasuvunakool2021highly}
Tunyasuvunakool,~K.; Adler,~J.; Wu,~Z.; Green,~T.; Zielinski,~M.;
  {\v{Z}}{\'{i}}dek,~A.; Bridgland,~A.; Cowie,~A.; Meyer,~C.; Laydon,~A.;
  Velankar,~S.; Kleywegt,~G.~J.; Bateman,~A.; Evans,~R.; Pritzel,~A.;
  Figurnov,~M.; Ronneberger,~O.; Bates,~R.; Kohl,~S. A.~A.; Potapenko,~A.;
  Ballard,~A.~J.; Romera-Paredes,~B.; Nikolov,~S.; Jain,~R.; Clancy,~E.;
  Reiman,~D.; Petersen,~S.; Senior,~A.~W.; Kavukcuoglu,~K.; Birney,~E.;
  Kohli,~P.; Jumper,~J.; Hassabis,~D. {Highly accurate protein structure
  prediction for the human proteome}. \emph{Nature} \textbf{2021}, \emph{596},
  590--596\relax
\mciteBstWouldAddEndPuncttrue
\mciteSetBstMidEndSepPunct{\mcitedefaultmidpunct}
{\mcitedefaultendpunct}{\mcitedefaultseppunct}\relax
\EndOfBibitem
\bibitem[Baek \latin{et~al.}(2021)Baek, DiMaio, Anishchenko, Dauparas,
  Ovchinnikov, Lee, Wang, Cong, Kinch, Schaeffer, Mill{\'{a}}n, Park, Adams,
  Glassman, DeGiovanni, Pereira, Rodrigues, van Dijk, Ebrecht, Opperman,
  Sagmeister, Buhlheller, Pavkov-Keller, Rathinaswamy, Dalwadi, Yip, Burke,
  Garcia, Grishin, Adams, Read, and Baker]{RoseTTAFold}
Baek,~M.; DiMaio,~F.; Anishchenko,~I.; Dauparas,~J.; Ovchinnikov,~S.;
  Lee,~G.~R.; Wang,~J.; Cong,~Q.; Kinch,~L.~N.; Schaeffer,~R.~D.;
  Mill{\'{a}}n,~C.; Park,~H.; Adams,~C.; Glassman,~C.~R.; DeGiovanni,~A.;
  Pereira,~J.~H.; Rodrigues,~A.~V.; van Dijk,~A.~A.; Ebrecht,~A.~C.;
  Opperman,~D.~J.; Sagmeister,~T.; Buhlheller,~C.; Pavkov-Keller,~T.;
  Rathinaswamy,~M.~K.; Dalwadi,~U.; Yip,~C.~K.; Burke,~J.~E.; Garcia,~K.~C.;
  Grishin,~N.~V.; Adams,~P.~D.; Read,~R.~J.; Baker,~D. {Accurate prediction of
  protein structures and interactions using a three-track neural network}.
  \emph{Science} \textbf{2021}, \emph{373}, 871--876\relax
\mciteBstWouldAddEndPuncttrue
\mciteSetBstMidEndSepPunct{\mcitedefaultmidpunct}
{\mcitedefaultendpunct}{\mcitedefaultseppunct}\relax
\EndOfBibitem
\bibitem[Webb \latin{et~al.}(2020)Webb, Jackson, Gil, and
  de~Pablo]{Webb2020Target}
Webb,~M.~A.; Jackson,~N.~E.; Gil,~P.~S.; de~Pablo,~J.~J. Targeted sequence
  design within the coarse-grained polymer genome. \emph{Sci. Adv.}
  \textbf{2020}, \emph{6}\relax
\mciteBstWouldAddEndPuncttrue
\mciteSetBstMidEndSepPunct{\mcitedefaultmidpunct}
{\mcitedefaultendpunct}{\mcitedefaultseppunct}\relax
\EndOfBibitem
\bibitem[Statt \latin{et~al.}(2020)Statt, Casademunt, Brangwynne, and
  Panagiotopoulos]{statt2020model}
Statt,~A.; Casademunt,~H.; Brangwynne,~C.~P.; Panagiotopoulos,~A.~Z. Model for
  disordered proteins with strongly sequence-dependent liquid phase behavior.
  \emph{J. Chem. Phys.} \textbf{2020}, \emph{152}, 075101\relax
\mciteBstWouldAddEndPuncttrue
\mciteSetBstMidEndSepPunct{\mcitedefaultmidpunct}
{\mcitedefaultendpunct}{\mcitedefaultseppunct}\relax
\EndOfBibitem
\bibitem[Statt \latin{et~al.}(2021)Statt, Kleeblatt, and
  Reinhart]{statt2021unsupervised}
Statt,~A.; Kleeblatt,~D.~C.; Reinhart,~W.~F. Unsupervised learning of
  sequence-specific aggregation behavior for a model copolymer. \emph{Soft
  Matter} \textbf{2021}, \emph{17}, 7697--7707\relax
\mciteBstWouldAddEndPuncttrue
\mciteSetBstMidEndSepPunct{\mcitedefaultmidpunct}
{\mcitedefaultendpunct}{\mcitedefaultseppunct}\relax
\EndOfBibitem
\bibitem[Meenakshisundaram \latin{et~al.}(2017)Meenakshisundaram, Hung, Patra,
  and Simmons]{Meenakshisundaram2017design}
Meenakshisundaram,~V.; Hung,~J.-H.; Patra,~T.~K.; Simmons,~D.~S. Designing
  Sequence-Specific Copolymer Compatibilizers Using a
  Molecular-Dynamics-Simulation-Based Genetic Algorithm. \emph{Macromolecules}
  \textbf{2017}, \emph{50}, 1155--1166\relax
\mciteBstWouldAddEndPuncttrue
\mciteSetBstMidEndSepPunct{\mcitedefaultmidpunct}
{\mcitedefaultendpunct}{\mcitedefaultseppunct}\relax
\EndOfBibitem
\bibitem[Ma \latin{et~al.}(2018)Ma, Huang, Zhang, and Luo]{ma2018determining}
Ma,~R.; Huang,~D.; Zhang,~T.; Luo,~T. Determining influential descriptors for
  polymer chain conformation based on empirical force-fields and molecular
  dynamics simulations. \emph{Chem. Phys. Lett.} \textbf{2018}, \emph{704},
  49--54\relax
\mciteBstWouldAddEndPuncttrue
\mciteSetBstMidEndSepPunct{\mcitedefaultmidpunct}
{\mcitedefaultendpunct}{\mcitedefaultseppunct}\relax
\EndOfBibitem
\bibitem[Arora \latin{et~al.}(2021)Arora, Lin, Rebello, Av-Ron, Mochigase, and
  Olsen]{arora2021random}
Arora,~A.; Lin,~T.-S.; Rebello,~N.~J.; Av-Ron,~S. H.~M.; Mochigase,~H.;
  Olsen,~B.~D. Random Forest Predictor for Diblock Copolymer Phase Behavior.
  \emph{ACS Macro Letters} \textbf{2021}, \emph{10}, 1339--1345\relax
\mciteBstWouldAddEndPuncttrue
\mciteSetBstMidEndSepPunct{\mcitedefaultmidpunct}
{\mcitedefaultendpunct}{\mcitedefaultseppunct}\relax
\EndOfBibitem
\bibitem[Patel \latin{et~al.}(2022)Patel, Borca, and
  Webb]{patel2022featurization}
Patel,~R.~A.; Borca,~C.~H.; Webb,~M.~A. Featurization strategies for polymer
  sequence or composition design by machine learning. \emph{Mol. Syst. Des.
  Eng.} \textbf{2022}, \emph{7}, 661--676\relax
\mciteBstWouldAddEndPuncttrue
\mciteSetBstMidEndSepPunct{\mcitedefaultmidpunct}
{\mcitedefaultendpunct}{\mcitedefaultseppunct}\relax
\EndOfBibitem
\bibitem[Ma \latin{et~al.}(2022)Ma, Zhang, and Luo]{ma2022exploring}
Ma,~R.; Zhang,~H.; Luo,~T. Exploring High Thermal Conductivity Amorphous
  Polymers Using Reinforcement Learning. \emph{ACS Applied Materials \&
  Interfaces} \textbf{2022}, \emph{14}, 15587--15598\relax
\mciteBstWouldAddEndPuncttrue
\mciteSetBstMidEndSepPunct{\mcitedefaultmidpunct}
{\mcitedefaultendpunct}{\mcitedefaultseppunct}\relax
\EndOfBibitem
\bibitem[Ma \latin{et~al.}(2019)Ma, Liu, Zhang, Liu, and Luo]{ma2019evaluating}
Ma,~R.; Liu,~Z.; Zhang,~Q.; Liu,~Z.; Luo,~T. Evaluating Polymer Representations
  via Quantifying Structure–Property Relationships. \emph{Journal of Chemical
  Information and Modeling} \textbf{2019}, \emph{59}, 3110--3119, PMID:
  31268306\relax
\mciteBstWouldAddEndPuncttrue
\mciteSetBstMidEndSepPunct{\mcitedefaultmidpunct}
{\mcitedefaultendpunct}{\mcitedefaultseppunct}\relax
\EndOfBibitem
\bibitem[Lin \latin{et~al.}(2019)Lin, Coley, Mochigase, Beech, Wang, Wang,
  Woods, Craig, Johnson, Kalow, Jensen, and Olsen]{lin2019bigsmiles}
Lin,~T.-S.; Coley,~C.~W.; Mochigase,~H.; Beech,~H.~K.; Wang,~W.; Wang,~Z.;
  Woods,~E.; Craig,~S.~L.; Johnson,~J.~A.; Kalow,~J.~A.; Jensen,~K.~F.;
  Olsen,~B.~D. BigSMILES: A Structurally-Based Line Notation for Describing
  Macromolecules. \emph{ACS Central Science} \textbf{2019}, \emph{5},
  1523--1531\relax
\mciteBstWouldAddEndPuncttrue
\mciteSetBstMidEndSepPunct{\mcitedefaultmidpunct}
{\mcitedefaultendpunct}{\mcitedefaultseppunct}\relax
\EndOfBibitem
\bibitem[Jindong and et~al.(2018)Jindong, and et~al.]{WangTLTutorial2018}
Jindong,~W.; et~al., Transfer Learning Tutorial. 2018;
  \url{https://github.com/jindongwang/transferlearning-tutorial}\relax
\mciteBstWouldAddEndPuncttrue
\mciteSetBstMidEndSepPunct{\mcitedefaultmidpunct}
{\mcitedefaultendpunct}{\mcitedefaultseppunct}\relax
\EndOfBibitem
\bibitem[Yang \latin{et~al.}(2020)Yang, Zhang, Dai, and
  Pan]{yang_zhang_dai_pan_2020}
Yang,~Q.; Zhang,~Y.; Dai,~W.; Pan,~S.~J. \emph{Transfer Learning}; Cambridge
  University Press, 2020\relax
\mciteBstWouldAddEndPuncttrue
\mciteSetBstMidEndSepPunct{\mcitedefaultmidpunct}
{\mcitedefaultendpunct}{\mcitedefaultseppunct}\relax
\EndOfBibitem
\bibitem[Yosinski \latin{et~al.}(2014)Yosinski, Clune, Bengio, and
  Lipson]{NIPS2014_375c7134}
Yosinski,~J.; Clune,~J.; Bengio,~Y.; Lipson,~H. How transferable are features
  in deep neural networks? Advances in Neural Information Processing Systems.
  2014; pp 1--9\relax
\mciteBstWouldAddEndPuncttrue
\mciteSetBstMidEndSepPunct{\mcitedefaultmidpunct}
{\mcitedefaultendpunct}{\mcitedefaultseppunct}\relax
\EndOfBibitem
\bibitem[Zhuang \latin{et~al.}(2020)Zhuang, Qi, Duan, Xi, Zhu, Zhu, Xiong, and
  He]{zhuang2020comprehensive}
Zhuang,~F.; Qi,~Z.; Duan,~K.; Xi,~D.; Zhu,~Y.; Zhu,~H.; Xiong,~H.; He,~Q. A
  comprehensive survey on transfer learning. \emph{Proceedings of the IEEE}
  \textbf{2020}, \emph{109}, 43--76\relax
\mciteBstWouldAddEndPuncttrue
\mciteSetBstMidEndSepPunct{\mcitedefaultmidpunct}
{\mcitedefaultendpunct}{\mcitedefaultseppunct}\relax
\EndOfBibitem
\bibitem[Pan and Yang(2009)Pan, and Yang]{pan2009survey}
Pan,~S.~J.; Yang,~Q. A survey on transfer learning. \emph{IEEE Transactions on
  Knowledge and Data Engineering} \textbf{2009}, \emph{22}, 1345--1359\relax
\mciteBstWouldAddEndPuncttrue
\mciteSetBstMidEndSepPunct{\mcitedefaultmidpunct}
{\mcitedefaultendpunct}{\mcitedefaultseppunct}\relax
\EndOfBibitem
\bibitem[Wang and Zheng(2015)Wang, and Zheng]{wang2015transfer}
Wang,~D.; Zheng,~T.~F. Transfer learning for speech and language processing.
  2015 Asia-Pacific Signal and Information Processing Association Annual Summit
  and Conference (APSIPA). 2015; pp 1225--1237\relax
\mciteBstWouldAddEndPuncttrue
\mciteSetBstMidEndSepPunct{\mcitedefaultmidpunct}
{\mcitedefaultendpunct}{\mcitedefaultseppunct}\relax
\EndOfBibitem
\bibitem[Kunze \latin{et~al.}(2017)Kunze, Kirsch, Kurenkov, Krug, Johannsmeier,
  and Stober]{kunze2017transfer}
Kunze,~J.; Kirsch,~L.; Kurenkov,~I.; Krug,~A.; Johannsmeier,~J.; Stober,~S.
  Transfer Learning for Speech Recognition on a Budget. \emph{CoRR}
  \textbf{2017}, \emph{abs/1706.00290}\relax
\mciteBstWouldAddEndPuncttrue
\mciteSetBstMidEndSepPunct{\mcitedefaultmidpunct}
{\mcitedefaultendpunct}{\mcitedefaultseppunct}\relax
\EndOfBibitem
\bibitem[Ng \latin{et~al.}(2015)Ng, Nguyen, Vonikakis, and Winkler]{ng2015deep}
Ng,~H.-W.; Nguyen,~V.~D.; Vonikakis,~V.; Winkler,~S. Deep Learning for Emotion
  Recognition on Small Datasets Using Transfer Learning. Proceedings of the
  2015 ACM on International Conference on Multimodal Interaction. New York, NY,
  USA, 2015; p 443–449\relax
\mciteBstWouldAddEndPuncttrue
\mciteSetBstMidEndSepPunct{\mcitedefaultmidpunct}
{\mcitedefaultendpunct}{\mcitedefaultseppunct}\relax
\EndOfBibitem
\bibitem[Yang \latin{et~al.}(2021)Yang, Zhang, Lv, and Wang]{yang2021image}
Yang,~X.; Zhang,~Y.; Lv,~W.; Wang,~D. Image recognition of wind turbine blade
  damage based on a deep learning model with transfer learning and an ensemble
  learning classifier. \emph{Renewable Energy} \textbf{2021}, \emph{163},
  386--397\relax
\mciteBstWouldAddEndPuncttrue
\mciteSetBstMidEndSepPunct{\mcitedefaultmidpunct}
{\mcitedefaultendpunct}{\mcitedefaultseppunct}\relax
\EndOfBibitem
\bibitem[Brown \latin{et~al.}(2020)Brown, Mann, Ryder, Subbiah, Kaplan,
  Dhariwal, Neelakantan, Shyam, Sastry, Askell, Agarwal, Herbert-Voss, Krueger,
  Henighan, Child, Ramesh, Ziegler, Wu, Winter, Hesse, Chen, Sigler, Litwin,
  Gray, Chess, Clark, Berner, McCandlish, Radford, Sutskever, and Amodei]{gpt3}
Brown,~T.; Mann,~B.; Ryder,~N.; Subbiah,~M.; Kaplan,~J.~D.; Dhariwal,~P.;
  Neelakantan,~A.; Shyam,~P.; Sastry,~G.; Askell,~A.; Agarwal,~S.;
  Herbert-Voss,~A.; Krueger,~G.; Henighan,~T.; Child,~R.; Ramesh,~A.;
  Ziegler,~D.; Wu,~J.; Winter,~C.; Hesse,~C.; Chen,~M.; Sigler,~E.; Litwin,~M.;
  Gray,~S.; Chess,~B.; Clark,~J.; Berner,~C.; McCandlish,~S.; Radford,~A.;
  Sutskever,~I.; Amodei,~D. Language Models are Few-Shot Learners. Advances in
  Neural Information Processing Systems. 2020; pp 1877--1901\relax
\mciteBstWouldAddEndPuncttrue
\mciteSetBstMidEndSepPunct{\mcitedefaultmidpunct}
{\mcitedefaultendpunct}{\mcitedefaultseppunct}\relax
\EndOfBibitem
\bibitem[Devlin \latin{et~al.}(2018)Devlin, Chang, Lee, and Toutanova]{bert}
Devlin,~J.; Chang,~M.-W.; Lee,~K.; Toutanova,~K. Bert: Pre-training of deep
  bidirectional transformers for language understanding. \emph{arXiv preprint
  arXiv:1810.04805} \textbf{2018}, \relax
\mciteBstWouldAddEndPunctfalse
\mciteSetBstMidEndSepPunct{\mcitedefaultmidpunct}
{}{\mcitedefaultseppunct}\relax
\EndOfBibitem
\bibitem[Tsubaki and Mizoguchi(2021)Tsubaki, and Mizoguchi]{Tsubaki2021quantum}
Tsubaki,~M.; Mizoguchi,~T. Quantum Deep Descriptor: Physically Informed
  Transfer Learning from Small Molecules to Polymers. \emph{Journal of Chemical
  Theory and Computation} \textbf{2021}, \emph{17}, 7814--7821, PMID:
  34846893\relax
\mciteBstWouldAddEndPuncttrue
\mciteSetBstMidEndSepPunct{\mcitedefaultmidpunct}
{\mcitedefaultendpunct}{\mcitedefaultseppunct}\relax
\EndOfBibitem
\bibitem[Sultan \latin{et~al.}(2018)Sultan, Wayment-Steele, and
  Pande]{Sultan2018transferable}
Sultan,~M.~M.; Wayment-Steele,~H.~K.; Pande,~V.~S. Transferable Neural Networks
  for Enhanced Sampling of Protein Dynamics. \emph{Journal of Chemical Theory
  and Computation} \textbf{2018}, \emph{14}, 1887--1894, PMID: 29529369\relax
\mciteBstWouldAddEndPuncttrue
\mciteSetBstMidEndSepPunct{\mcitedefaultmidpunct}
{\mcitedefaultendpunct}{\mcitedefaultseppunct}\relax
\EndOfBibitem
\bibitem[Käser \latin{et~al.}(2021)Käser, Boittier, Upadhyay, and
  Meuwly]{kaser2021transfer}
Käser,~S.; Boittier,~E.~D.; Upadhyay,~M.; Meuwly,~M. Transfer Learning to
  CCSD(T): Accurate Anharmonic Frequencies from Machine Learning Models.
  \emph{Journal of Chemical Theory and Computation} \textbf{2021}, \emph{17},
  3687--3699, PMID: 33960787\relax
\mciteBstWouldAddEndPuncttrue
\mciteSetBstMidEndSepPunct{\mcitedefaultmidpunct}
{\mcitedefaultendpunct}{\mcitedefaultseppunct}\relax
\EndOfBibitem
\bibitem[Ma \latin{et~al.}(2020)Ma, Colón, and Luo]{ma2020transfer}
Ma,~R.; Colón,~Y.~J.; Luo,~T. Transfer Learning Study of Gas Adsorption in
  Metal–Organic Frameworks. \emph{ACS Appl. Mater. Interfaces} \textbf{2020},
  \emph{12}, 34041--34048, PMID: 32613831\relax
\mciteBstWouldAddEndPuncttrue
\mciteSetBstMidEndSepPunct{\mcitedefaultmidpunct}
{\mcitedefaultendpunct}{\mcitedefaultseppunct}\relax
\EndOfBibitem
\bibitem[Liu \latin{et~al.}(2020)Liu, Jiang, and Luo]{liu2020leverage}
Liu,~Z.; Jiang,~M.; Luo,~T. Leverage electron properties to predict phonon
  properties via transfer learning for semiconductors. \emph{Science Advances}
  \textbf{2020}, \emph{6}, eabd1356\relax
\mciteBstWouldAddEndPuncttrue
\mciteSetBstMidEndSepPunct{\mcitedefaultmidpunct}
{\mcitedefaultendpunct}{\mcitedefaultseppunct}\relax
\EndOfBibitem
\bibitem[Wu \latin{et~al.}(2019)Wu, Kondo, Kakimoto, Yang, Yamada, Kuwajima,
  Lambard, Hongo, Xu, Shiomi, Schick, Morikawa, and Yoshida]{wu2019machine}
Wu,~S.; Kondo,~Y.; Kakimoto,~M.-a.; Yang,~B.; Yamada,~H.; Kuwajima,~I.;
  Lambard,~G.; Hongo,~K.; Xu,~Y.; Shiomi,~J.; Schick,~C.; Morikawa,~J.;
  Yoshida,~R. {Machine-learning-assisted discovery of polymers with high
  thermal conductivity using a molecular design algorithm}. \emph{npj
  Computational Materials} \textbf{2019}, \emph{5}, 66\relax
\mciteBstWouldAddEndPuncttrue
\mciteSetBstMidEndSepPunct{\mcitedefaultmidpunct}
{\mcitedefaultendpunct}{\mcitedefaultseppunct}\relax
\EndOfBibitem
\bibitem[Thompson \latin{et~al.}(2022)Thompson, Aktulga, Berger, Bolintineanu,
  Brown, Crozier, in~'t Veld, Kohlmeyer, Moore, Nguyen, Shan, Stevens,
  Tranchida, Trott, and Plimpton]{LAMMPS}
Thompson,~A.~P.; Aktulga,~H.~M.; Berger,~R.; Bolintineanu,~D.~S.; Brown,~W.~M.;
  Crozier,~P.~S.; in~'t Veld,~P.~J.; Kohlmeyer,~A.; Moore,~S.~G.;
  Nguyen,~T.~D.; Shan,~R.; Stevens,~M.~J.; Tranchida,~J.; Trott,~C.;
  Plimpton,~S.~J. {LAMMPS} - a flexible simulation tool for particle-based
  materials modeling at the atomic, meso, and continuum scales. \emph{Comp.
  Phys. Comm.} \textbf{2022}, \emph{271}, 108171\relax
\mciteBstWouldAddEndPuncttrue
\mciteSetBstMidEndSepPunct{\mcitedefaultmidpunct}
{\mcitedefaultendpunct}{\mcitedefaultseppunct}\relax
\EndOfBibitem
\bibitem[Darve \latin{et~al.}(2008)Darve, Rodr\'{i}guez-G\'{o}mez, and
  Pohorille]{Darve2008adaptive}
Darve,~E.; Rodr\'{i}guez-G\'{o}mez,~D.; Pohorille,~A. Adaptive biasing force
  method for scalar and vector free energy calculations. \emph{J. Chem. Phys.}
  \textbf{2008}, \emph{128}, 144120\relax
\mciteBstWouldAddEndPuncttrue
\mciteSetBstMidEndSepPunct{\mcitedefaultmidpunct}
{\mcitedefaultendpunct}{\mcitedefaultseppunct}\relax
\EndOfBibitem
\bibitem[Sidky \latin{et~al.}(2018)Sidky, Col{\'{o}}n, Helfferich, Sikora,
  Bezik, Chu, Giberti, Guo, Jiang, Lequieu, Li, Moller, Quevillon, Rahimi,
  Ramezani-Dakhel, Rathee, Reid, Sevgen, Thapar, Webb, Whitmer, and
  de~Pablo]{SSAGES}
Sidky,~H.; Col{\'{o}}n,~Y.~J.; Helfferich,~J.; Sikora,~B.~J.; Bezik,~C.;
  Chu,~W.; Giberti,~F.; Guo,~A.~Z.; Jiang,~X.; Lequieu,~J.; Li,~J.; Moller,~J.;
  Quevillon,~M.~J.; Rahimi,~M.; Ramezani-Dakhel,~H.; Rathee,~V.~S.;
  Reid,~D.~R.; Sevgen,~E.; Thapar,~V.; Webb,~M.~A.; Whitmer,~J.~K.;
  de~Pablo,~J.~J. {SSAGES: Software Suite for Advanced General Ensemble
  Simulations}. \emph{The Journal of Chemical Physics} \textbf{2018},
  \emph{148}, 044104\relax
\mciteBstWouldAddEndPuncttrue
\mciteSetBstMidEndSepPunct{\mcitedefaultmidpunct}
{\mcitedefaultendpunct}{\mcitedefaultseppunct}\relax
\EndOfBibitem
\bibitem[Shi \latin{et~al.}(2022)Shi, Huang, Gygi, and Whitmer]{shi2022free}
Shi,~J.; Huang,~S.; Gygi,~F.; Whitmer,~J.~K. Free-Energy Landscape and
  Isomerization Rates of Au4 Clusters at Finite Temperatures. \emph{The Journal
  of Physical Chemistry A} \textbf{2022}, \emph{126}, 3392--3400, PMID:
  35584205\relax
\mciteBstWouldAddEndPuncttrue
\mciteSetBstMidEndSepPunct{\mcitedefaultmidpunct}
{\mcitedefaultendpunct}{\mcitedefaultseppunct}\relax
\EndOfBibitem
\bibitem[Shi \latin{et~al.}(2020)Shi, Sidky, and Whitmer]{shi2020automated}
Shi,~J.; Sidky,~H.; Whitmer,~J.~K. Automated determination of n-cyanobiphenyl
  and n-cyanobiphenyl binary mixtures elastic constants in the nematic phase
  from molecular simulation. \emph{Mol. Syst. Des. Eng.} \textbf{2020},
  \emph{5}, 1131--1136\relax
\mciteBstWouldAddEndPuncttrue
\mciteSetBstMidEndSepPunct{\mcitedefaultmidpunct}
{\mcitedefaultendpunct}{\mcitedefaultseppunct}\relax
\EndOfBibitem
\bibitem[Leonhard and Whitmer(2019)Leonhard, and Whitmer]{leonhard2019accurate}
Leonhard,~A.~C.; Whitmer,~J.~K. Accurate Determination of Cavitand Binding Free
  Energies via Unrestrained Advanced Sampling. \emph{Journal of Chemical Theory
  and Computation} \textbf{2019}, \emph{15}, 5761--5768, PMID: 31566977\relax
\mciteBstWouldAddEndPuncttrue
\mciteSetBstMidEndSepPunct{\mcitedefaultmidpunct}
{\mcitedefaultendpunct}{\mcitedefaultseppunct}\relax
\EndOfBibitem
\bibitem[Cortés-Morales \latin{et~al.}(2021)Cortés-Morales, Rathee, Ghobadi,
  and Whitmer]{cortes2021a}
Cortés-Morales,~E.~C.; Rathee,~V.~S.; Ghobadi,~A.; Whitmer,~J.~K. A molecular
  view of plasticization of polyvinyl alcohol. \emph{The Journal of Chemical
  Physics} \textbf{2021}, \emph{155}, 174903\relax
\mciteBstWouldAddEndPuncttrue
\mciteSetBstMidEndSepPunct{\mcitedefaultmidpunct}
{\mcitedefaultendpunct}{\mcitedefaultseppunct}\relax
\EndOfBibitem
\bibitem[Zhang \latin{et~al.}(2021)Zhang, Lipton, Li, and Smola]{zhang2021dive}
Zhang,~A.; Lipton,~Z.~C.; Li,~M.; Smola,~A.~J. Dive into Deep Learning.
  \emph{arXiv preprint arXiv:2106.11342} \textbf{2021}, \relax
\mciteBstWouldAddEndPunctfalse
\mciteSetBstMidEndSepPunct{\mcitedefaultmidpunct}
{}{\mcitedefaultseppunct}\relax
\EndOfBibitem
\bibitem[Xu \latin{et~al.}(2015)Xu, Wang, Chen, and Li]{xu2015empirical}
Xu,~B.; Wang,~N.; Chen,~T.; Li,~M. Empirical evaluation of rectified
  activations in convolutional network. \emph{arXiv preprint arXiv:1505.00853}
  \textbf{2015}, \relax
\mciteBstWouldAddEndPunctfalse
\mciteSetBstMidEndSepPunct{\mcitedefaultmidpunct}
{}{\mcitedefaultseppunct}\relax
\EndOfBibitem
\bibitem[Kingma and Ba(2014)Kingma, and Ba]{kingma2014adam}
Kingma,~D.~P.; Ba,~J. Adam: A method for stochastic optimization. \emph{arXiv
  preprint arXiv:1412.6980} \textbf{2014}, \relax
\mciteBstWouldAddEndPunctfalse
\mciteSetBstMidEndSepPunct{\mcitedefaultmidpunct}
{}{\mcitedefaultseppunct}\relax
\EndOfBibitem
\bibitem[Paszke \latin{et~al.}(2019)Paszke, Gross, Massa, Lerer, Bradbury,
  Chanan, Killeen, Lin, Gimelshein, Antiga, Desmaison, Kopf, Yang, DeVito,
  Raison, Tejani, Chilamkurthy, Steiner, Fang, Bai, and Chintala]{pytorch}
Paszke,~A.; Gross,~S.; Massa,~F.; Lerer,~A.; Bradbury,~J.; Chanan,~G.;
  Killeen,~T.; Lin,~Z.; Gimelshein,~N.; Antiga,~L.; Desmaison,~A.; Kopf,~A.;
  Yang,~E.; DeVito,~Z.; Raison,~M.; Tejani,~A.; Chilamkurthy,~S.; Steiner,~B.;
  Fang,~L.; Bai,~J.; Chintala,~S. In \emph{Advances in Neural Information
  Processing Systems 32}; Wallach,~H., Larochelle,~H., Beygelzimer,~A.,
  d'Alch\'{e} Buc,~F., Fox,~E., Garnett,~R., Eds.; Curran Associates, Inc.,
  2019; pp 8024--8035\relax
\mciteBstWouldAddEndPuncttrue
\mciteSetBstMidEndSepPunct{\mcitedefaultmidpunct}
{\mcitedefaultendpunct}{\mcitedefaultseppunct}\relax
\EndOfBibitem
\bibitem[Altmann \latin{et~al.}(2010)Altmann, Toloşi, Sander, and
  Lengauer]{altmann2010permutation}
Altmann,~A.; Toloşi,~L.; Sander,~O.; Lengauer,~T. {Permutation importance: a
  corrected feature importance measure}. \emph{Bioinformatics} \textbf{2010},
  \emph{26}, 1340--1347\relax
\mciteBstWouldAddEndPuncttrue
\mciteSetBstMidEndSepPunct{\mcitedefaultmidpunct}
{\mcitedefaultendpunct}{\mcitedefaultseppunct}\relax
\EndOfBibitem
\bibitem[TeamHG-Memex(2020)]{ELI5}
TeamHG-Memex, ELI5. \url{https://github.com/TeamHG-Memex/eli5}, 2020\relax
\mciteBstWouldAddEndPuncttrue
\mciteSetBstMidEndSepPunct{\mcitedefaultmidpunct}
{\mcitedefaultendpunct}{\mcitedefaultseppunct}\relax
\EndOfBibitem
\bibitem[Redmon \latin{et~al.}(2016)Redmon, Divvala, Girshick, and
  Farhadi]{redmon2016you}
Redmon,~J.; Divvala,~S.; Girshick,~R.; Farhadi,~A. You only look once: Unified,
  real-time object detection. Proceedings of the IEEE conference on computer
  vision and pattern recognition. 2016; pp 779--788\relax
\mciteBstWouldAddEndPuncttrue
\mciteSetBstMidEndSepPunct{\mcitedefaultmidpunct}
{\mcitedefaultendpunct}{\mcitedefaultseppunct}\relax
\EndOfBibitem
\end{mcitethebibliography}

\end{document}